\documentclass[apj]{emulateapj}
\usepackage[utf8]{inputenc}
\usepackage[T1]{fontenc}
\usepackage{graphicx}
\usepackage{captcont}
\usepackage[english]{babel}
\usepackage{amssymb}
\usepackage{float}
\usepackage{multirow}
\usepackage{color}
\usepackage{threeparttable}
\usepackage{color}
\usepackage{arydshln}

\definecolor{ashgrey}{rgb}{0.7, 0.75, 0.71}
\definecolor{darkgray}{rgb}{0.66, 0.66, 0.66}
\definecolor{trolleygrey}{rgb}{0.5, 0.5, 0.5}
 	\definecolor{gray}{rgb}{0.5, 0.5, 0.5}
\definecolor{davysgrey}{rgb}{0.33, 0.33, 0.33}
 	
\usepackage[normalem]{ulem}

\usepackage{multirow}
\usepackage{xcolor}

\usepackage{float}
\usepackage{times}

\slugcomment{}

\shorttitle{Chaos in the Early Solar System}
\shortauthors{Izidoro et al.}

\begin{document} 


\title{The Asteroid Belt as a Relic from a Chaotic  Early Solar System}

\author{André Izidoro}
\affil{ Laboratoire d'astrophysique de Bordeaux, Univ. Bordeaux, CNRS, B18N, allée Geoffroy Saint-Hilaire, 33615 Pessac, France \\ Capes Foundation, Ministry of Education of Brazil, Brasília/DF 70040-020, Brazil.}
\email{izidoro.costa@gmail.com}

\author{Sean N. Raymond}
\affil{ Laboratoire d'astrophysique de Bordeaux, Univ. Bordeaux, CNRS, B18N, allée Geoffroy Saint-Hilaire, 33615 Pessac, France}

\author{Arnaud Pierens}
\affil{ Laboratoire d'astrophysique de Bordeaux, Univ. Bordeaux, CNRS, B18N, allée Geoffroy Saint-Hilaire, 33615 Pessac, France}

\author{Alessandro Morbidelli }
\affil{University of Nice-Sophia Antipolis, CNRS, Observatoire de la Côte d’Azur, Laboratoire Lagrange, BP 4229, 06304 Nice Cedex 4, France.}

\author{Othon C. Winter}
\affil{UNESP, Univ. Estadual Paulista - Grupo de Dinâmica Orbital \& Planetologia, Guaratinguetá, CEP 12.516-410, São Paulo, Brazil}

\author{David Nesvorn{\`y}}
\affil{Department of Space Studies, Southwest Research Institute, 1050 Walnut St., Suite 300, Boulder, CO 80302, USA}


\begin{abstract}

The orbital structure of the asteroid belt holds a record of the Solar System's dynamical history. The current belt only contains ${\rm \sim 10^{-3}}$ Earth masses yet the asteroids' orbits are dynamically excited, with a large spread in eccentricity and inclination. In the context of models of terrestrial planet formation, the belt may have been excited by Jupiter's orbital migration. The terrestrial planets can also be reproduced without invoking a migrating Jupiter; however, as it requires a severe mass deficit beyond Earth's orbit, this model systematically under-excites the asteroid belt.  Here we show that the orbits of the asteroids may have been excited to their current state if Jupiter and Saturn's early orbits were chaotic.  Stochastic variations in the gas giants' orbits cause resonances to continually jump across the main belt and excite the asteroids' orbits on a timescale of tens of millions of years. While hydrodynamical simulations show that the gas giants were likely in mean motion resonance at the end of the gaseous disk phase, small perturbations could have driven them into a chaotic but stable state. The gas giants' current orbits were achieved later, during an instability in the outer Solar System. Although it is well known that the present-day Solar System exhibits chaotic behavior, our results suggest that the early Solar System may also have been chaotic.  

\end{abstract}

\section{Introduction}
The distribution of asteroids strongly constrains planet formation models. While the terrestrial planets' orbits are nearly circular and coplanar, the orbital eccentricities of asteroids are excited, filling parameter space from $e=0$ to 0.3, and inclination $i=0$ to $20^\circ$. The asteroid belt total mass is also only $\sim 10^{-3}$ Earth masses.

There are two basic views on how the inner solar system was built, with different implications for the asteroid belt.  In one view, the asteroid belt contained a few Earth masses in solid material but was rapidly depleted and excited by  dynamical mechanisms. Gravitational scattering of asteroids by a population of Moon- to Mars-sized planetary embryos originally in the belt can promote significant depletion and excitation of the belt (\cite{wetherilletal78,wetherill92a, agnoretal99,whetherill86,chamberswetherill98,petitetal99,chambers01,petitetal01,petitetal02,obrienetal07}). One problem with this scenario in the context of terrestrial planet formation is that Mars analogs produced in these simulations are far more massive than the actual one (e.g., \cite{raymondetal04,raymondetal06,morishimaetal08,raymondetal09,obrienetal06,izidoroetal13,lykawkaito13,izidoroetal14a,fischerciesla14}). The Grand Tack scenario \citep{walshetal11} -- which invokes the inward-then outward migration of Jupiter through the asteroid belt region -- removes enough mass beyond 1 AU to explain why Mars is much smaller than Earth and to sculpt the asteroid belt in a way that will become consistent with its current structure via subsequent dynamical evolution \citep{roignesvorny15,deiennoetal16}. In the opposite view, the asteroid belt was low-mass even at early times \citep{izidoroetal15b,levisonetal15,moriartyfischer15,drazkowskaetal16}, and Jupiter and Saturn did not migrate across the asteroid belt. In this framework, a primordial low mass asteroid belt should be far less dynamically excited than the observed one \citep{izidoroetal15b}, and what remains to be explained is the belt's dynamical excitation (for a recent review see \cite{morbidellietal15}).  

 In this paper we propose a novel mechanism for explaining the dynamical excitation of the asteroid belt.  The mechanism relies on the chaotic evolution of Jupiter and Saturn's orbits at early times. In section 2 we present an example for the origin of chaos in Jupiter and Saturn's early orbits. In section 3 we present our results for the chaotic excitation of the asteroid belt. In section 4 we more fully address the possible origins of chaos in the giant planets' orbits, and present alternative scenarios to trigger chaos. In section 5 we discuss on the implication of our results for models of solar system formation.  Finally, in section 6 we briefly summarize our findings.

\section{An Example of Chaos in Jupiter and Saturn's early orbits}

The excitation of the asteroid belt took place after the gaseous protoplanetary disk had dissipated, yet it is during the disk phase that the gas giants' orbits could have changed most dramatically due to orbital migration. Embedded in the gaseous disk, Jupiter and Saturn systematically migrate into mean motion resonance (MMR), where their orbital periods are related by a ratio of small integers \citep{massetsnellgrove01,morbidellicrida07,pierensnelson08,dangelomarzari12,pierensetal14}. The most common are the 3:2 and 2:1 MMRs. The present-day orbits of the giant planets were achieved later, after the gaseous disk was gone, during a dynamical instability \citep{nesvornymorbidelli12}.  In our model, the asteroids were excited between the dissipation of the disk and the instability.

In hydrodynamical simulations Jupiter and Saturn's migration typically leads to deep capture in resonance, with orbits characterized by regular motion. However, very small perturbations may push them into chaos \citep{sandorkley06,batyginmorbidelli13}. Perturbations come from (i) dispersal of the gaseous disk and the corresponding loss of its damping \citep{papaloizoularwood00,cresswellnelson08}; (ii) gravitational forcing from the ice giants, both during their inward migration \citep{izidoroetal15a} and after the dissipation of the gaseous disk; and (iii) gravitational scattering of small remnant planetary embryos and planetesimals in the giant planet region. 

We performed a suite of numerical experiments to show that seemingly trivial perturbations can trigger chaos in the gas giants' orbits.  Figure 1 shows one simulation in which a regularly-evolving configuration of Jupiter and Saturn (the JSREG simulation), locked in 2:1 resonance, became chaotic as a result of the ejection of a Mars-sized embryo from the system. To perform this simulation we used the Mercury integrator \citep{chambers99}. The system was composed by the central solar mass star, the fully formed Jupiter and Saturn in the 2:1 MMR, and a Mars-mass planetary embryo. Jupiter was initially at 5.25 AU, Saturn at $\sim$8.33 AU, and the planetary embryo at 12 AU. The gas giants eccentricities were initially 0.025 and their mutual orbital inclination 0.5 degrees. The planetary embryo started with zero orbital eccentricity and inclination. We used an integration timestep of 100 days and assumed that the gaseous disk was already fully dissipated. 

While the planetary embryo in Fig. 1 was only 1/4000-th the combined mass of the gas giants, it triggered chaos. The gas giants' orbits were chaotic but dynamically stable for long timescales, consistent with a late dynamical instability that re-arranged their orbits \citep{tsiganisetal05,morbidellietal07,levisonetal11}.  The perturbed giant planets even remained in 2:1 resonance with modest (but chaotic) eccentricities and inclinations ( see for example Fig. 2). Depending on the nature and strength of the perturbation, chaos was generated in up to 100\% of our simulations starting from a regular, resonant configuration (see also section 3.2). In fact, stochastic forcing from turbulent density fluctuations during the disk phase may have pushed the planets out of (deep) resonance \citep{adamsetal08,lecoanetetal09, pierensetal11}, onto nearby orbits where the density of chaos is high (see Appendix).

\begin{figure}[h]
\centering
\includegraphics[scale=2.71]{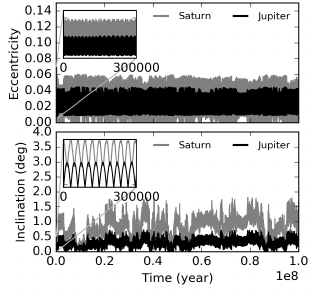}
\caption{The onset of chaos in a characteristic dynamical simulation. Jupiter and Saturn started in the JSREG configuration, locked in 2:1 MMR with low eccentricities (starting eccentricities of 0.025 for both planets) and exhibiting regular motion. The evolution of their orbital eccentricities/inclinations are shown in the top/bottom panel.  Gravitational scattering of a Mars-mass embryo initially at 12~AU triggered chaos in the giant planets' orbits. The planets remained in resonance, as confirmed by libration of the 2:1 MMR critical angle $2\lambda_{Sat} - \lambda_{Jup} - \varpi_{Jup}$, on chaotic but long-term stable orbits. $\lambda_{Sat}$, $\lambda_{Jup}$, and  $\varpi_{Jup}$ are Saturn's mean longitude, Jupiter's mean longitude and longitude of pericenter, respectively. The initial conditions of this simulation are provided in the Appendix.}
\end{figure}

This simulation (Fig. 1) represents a simple proof of concept; a regular orbital configuration of Jupiter and Saturn can easily be converted into a chaotic one.  The perturbation required must be strong enough to transition the system to a new dynamical state but not so strong as to make the system dynamically unstable. The magnitude of the perturbation in the simulation from Fig 1 is entirely plausible, as it is likely that some leftovers remained when the protoplanetary disk dissipated.  Indeed, the so-called ``late veneer'' represents geochemical evidence that planetary leftovers remained scattered throughout the inner Solar System after the Moon-forming impact on Earth, long after the dissipation of the disk (e.g., \cite{dayetal07,walker09,bottkeetal10,jacobsonetal14}). In section 4 we present additional scenarios for generating chaos in Jupiter and Saturn's early orbits.

\section{Chaotic Excitation of the Asteroid Belt}

We now turn our attention to showing how the asteroids' orbits may have been excited by chaos in the orbits of Jupiter and Saturn.  We used the Symba \citep{duncanetal98} and Swift \citep{levisonduncan94}, Mercury \citep{chambers99} and Rebound \citep{reinspiegel14,reintamayo15} integrators  to perform our simulations of the belt excitation. Asteroids were modeled as massless test particles. The integration timestep in all our simulations was at most 1/20th the orbital period of the innermost body in the system. We stress that we are aware of an issue with Symba5 which compromises the performance of the integrator in the case where planets have close-encounters every orbital period. When this kind of evolution takes place it degrades the symplectic nature of the integrator causing large errors. This may be the case for example when Jupiter and Saturn  evolve in a compact resonant configuration (e.g. 3:2 mean motion resonance). To make sure the chaos observed in our simulations have no numeric origin we have tested an ample number of integrators over  different simulations presented here. In Rebound we test both the WHFAST and IAS15 integrators and in Mercury we performed tests with the Bulirsch–Stoer and ``Hybrid'' integrators. 

Figure 2 shows the dynamical evolution of Jupiter and Saturn in the JSREG and JSCHA simulations, which we use as fiducial cases to illustrate and contrast mechanisms of excitation of the asteroid belt.  These two simulations were generated from almost identical initial orbital arrangements; the only difference is that Jupiter and Saturn's eccentricities were each set to 0.025 in the JSREG simulation (resulting in regular motion) and to 0.03 in the JSCHA simulation (which triggered chaos). In both simulations, Jupiter and Saturn's semi-major axes are initially 5.4 and  about $\sim$8.57 AU, respectively. Their mutual initial orbital inclinations are 0.5 degrees. Their argument of pericenter and longitude of the ascending node are set zero. Jupiter's mean anomaly is initially zero and Saturn's mean anomaly is initially 180 degrees.

Figure 2 shows the evolution of the gas giant's period ratio, eccentricities, orbital inclinations, and an angle associated with the 2:1 mean motion resonance between the planets. In both simulations the planets are in 2:1 resonance since the critical angle $ \phi_2 = 2\lambda_{Sat} -\lambda_{Jup} - \varpi_{Jup}$ librates around zero degree, where $\lambda_{Jup}$ and $\lambda_{Sat}$ are the mean longitudes of Jupiter and Saturn, respectively. The critical angle $ \phi_1 = 2\lambda_{Sat} -\lambda_{Jup} - \varpi_{Sat}$ circulates, where $\varpi_{Sat}$ is Saturn's longitude of pericenter.  Thus, the planets are not in apsidal corotation resonance -- in which the planets undergo apsidal libration as well as libration of both resonant arguments \citep{michtchenkoetal08} -- as shown by the circulation of the angles $\varpi_{Sat} - \varpi_{Jup}$ (Figure 2, bottom panels). Yet the eccentricity and inclination evolution of Jupiter and Saturn are quite different in the the two simulations.  In particular, there are larger variations of eccentricity and inclination over 100 Myr in the chaotic case. These variations are linked with precession of longitudes which translate to shifting resonances and gravitational perturbations in the belt.

\begin{figure*}[h]
\centering
\includegraphics[scale=2.61]{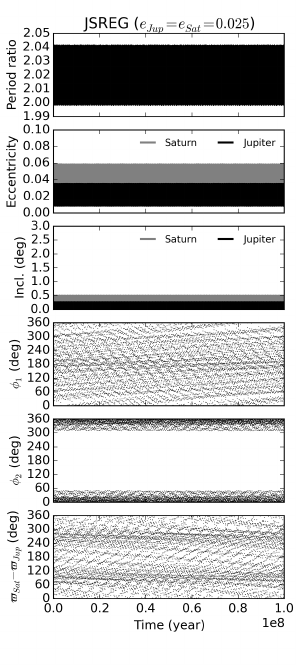}
\includegraphics[scale=2.61]{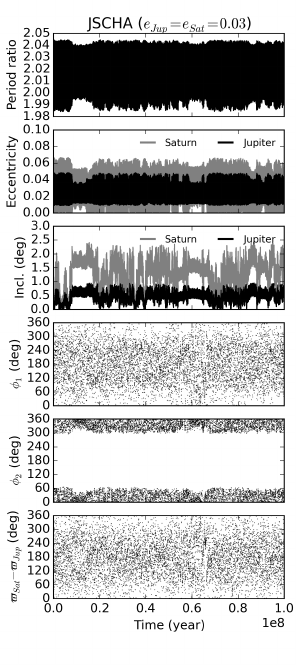}
\caption{Dynamical evolution of regular Jupiter and Saturn in the JSREG and JSCHA simulations. In both cases the planets are in the 2:1 mean motion resonance (the critical angle $\phi_2 = 2\lambda_{Sat} - \lambda_{Jup} - \varpi_{Jup}$ associated with the 2:1 mean motion resonance librate around zero degree while the $\phi_1 = 2\lambda_{Sat} - \lambda_{Jup} - \varpi_{Sat}$ circulates). $\lambda_{Sat}$ and  $\varpi_{Sat}$ are Saturn's mean longitude and longitude of pericenter. $\lambda_{Jup}$, and $\varpi_{Jup}$  are Jupiter's mean longitude and longitude of pericenter. The initial conditions of these two simulations are provided in the Appendix }
\end{figure*}

We now show how, once the gaseous disk was gone, chaos in Jupiter and Saturn's early orbits may have excited the asteroids' orbits even if the belt's primordial mass was very low (comparable to its current mass).  The asteroid belt is speckled with resonances, locations where there is an integer match between characteristic orbital frequencies of asteroids and the giant planets. MMRs are located where an asteroid's orbital period forms an integer ratio with a planet's period (Jupiter in this case). In secular resonances (SRs), a quantity related to the precession of an asteroid's orbit matches one of the giant planets. The most important SRs in the main belt are the $\nu_{6}$ and $\nu_{16}$ resonances, where an asteroid's apsidal and nodal precession rate (or frequency), respectively, match that of Saturn  \citep{froeschlescholl89,morbidellihenrard91}. When Jupiter and Saturn's orbits evolve in a regular fashion, MMRs and SRs are stationary, so asteroids in certain parts of the belt are excited whereas asteroids in other parts are not  \citep{morbidellihenrard91}.

When the giant planets evolve chaotically, their orbital alignments undergo stochastic jumps, i.e, they may precess at many different frequencies and their orbits may even have its direction of apsidal precession temporarily reversed. The location of SRs within the belt undergo corresponding jumps (MMRs are much less sensitive to such variations). When a resonance jumps to the location of a given asteroid its orbit is significantly excited on a relatively short ($\sim 10^{4-6}$~year) timescale. Figure 3 shows how the $\nu_6$ and $\nu_{16}$ SRs pump the eccentricity and orbital inclinations, respectively, of two asteroids.  Note that the time intervals shown in both plots of Figure 3 were purposely chosen to clearly illustrate the effect of the respective resonances. The lifetime of a particle in the belt  depends on the chaotic evolution of the gas giants and the particle's initial orbit. We discuss and show  in the Appendix that the effects of the chaotic excitation can also eject particles from the system and even empty  parts of the belt.

In addition to the $\nu_6$ and $\nu_{16}$ other resonances also play a role in pumping asteroids' eccentricities and/or inclinations (see Appendix). These include resonances resulting from the linear combination of principal secular frequencies (nonlinear secular resonances), mean motion resonances, secondary resonances (linear combination between the main secular and short period frequencies associated with orbital period) and Kozai resonances. Another factor is that, unlike MMRs which are linked to a given orbital radius, the locations of SRs are a function of an asteroid's proper inclination and eccentricity \citep{froeschlescholl89,morbidellihenrard91}.

\begin{figure*}
\centering
\includegraphics[scale=1.4]{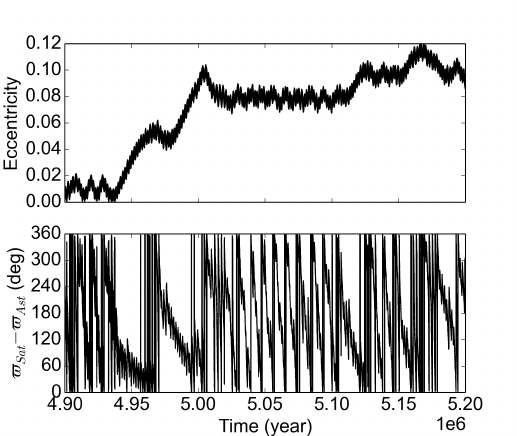} 
\includegraphics[scale=1.4]{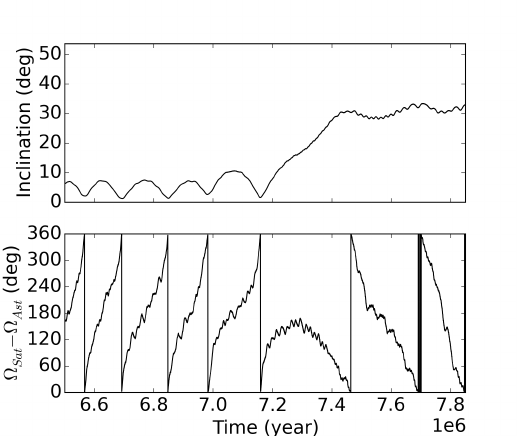}  
\caption{Chaotic excitation of the orbits of two asteroids (modeled as massless test particles) in a dynamical simulation.  {\bf Left:} The eccentricity of an asteroid at 3 AU increases rapidly during a short interval while it is locked in the $\nu_6$ secular resonance. The $\nu_{6}$ occurs when the angle ${\rm \varpi_{Sat}-\varpi_{Ast}}$ librates, where ${\rm \varpi_{Sat}}$ and ${\rm \varpi_{Ast}}$ are the longitudes of pericenter of Saturn and the asteroid, respectively.  {\bf Right:} The inclination of an asteroid at 2.7 AU is pumped during temporary capture in the $\nu_{16}$ secular resonance. The $\nu_{16}$ occurs when the angle ${\rm \Omega_{Sat}-\Omega_{Ast}}$ librates, where ${\rm \Omega_{Sat}}$ and ${\rm \Omega_{Ast}}$ are the longitudes of ascending node of Saturn and the asteroid, respectively. The asteroids' semi-major axes remain roughly constant. 
}
\end{figure*}

We can understand the chaotic excitation of asteroids using a Fourier analysis of Saturn's longitude of pericenter $\varpi_{Sat}$ (Fig. 4; top panels).  When Jupiter and Saturn undergo regular motion, the power spectrum of $\varpi_{Sat}$ is peaked, dominated by characteristic frequencies (in particular the ${\mathrm g}_6$ frequency at $\sim$1/62000 ${\rm yr}^{-1}$, which controls the location of the $\nu_6$ SR) and their harmonics \citep{laskar90,laskar93,michtchenkoferraz-mello95}.  The peaked nature of the power spectrum indicates that Saturn's precession frequency -- and thus the location of the $\nu_6$ and other resonances -- is fixed. This explains why eccentricities and inclinations of asteroids are only strongly excited at specific locations in the belt (Fig. 4).  In contrast, the power spectrum of a simulation in which Jupiter and Saturn's orbits evolve chaotically (the JSCHA simulation) shows a broad band of frequencies instead of a few strong peaks. In this case the SR $\nu_6$ jumps across the entire belt because $\varpi_{Sat}$ precesses at many different frequencies due to Saturn's chaotic interactions with Jupiter. Note from the frequency analysis in Fig. 4 that in the JSREG simulation the longitude of pericenter of Saturn precesses (slowly) positively  while the longitude of pericenter of Jupiter precesses (quickly) backwards. If planets were in apsidal corotation that would imply both their longitudes of pericenter would precess in the same direction (negative). Because asteroids suffering secular perturbation precesses always positively, the $\nu_6$ could not exist in the belt if asteroids' and Saturn's longitude of pericenter precess in different directions.

\begin{figure*}
\centering
\includegraphics[scale=2.88]{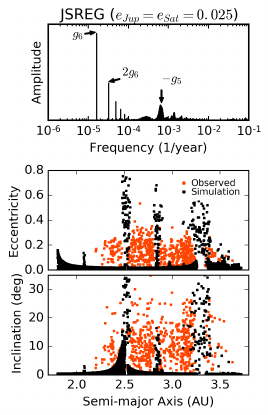}
\includegraphics[scale=2.88]{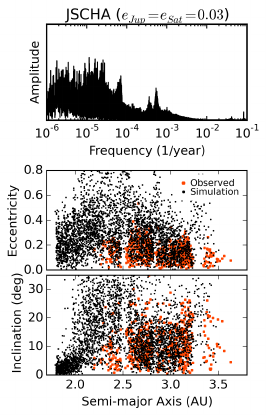}
\caption{\small Fourier analysis and excitation of the asteroid belt by the gas giants in two N-body simulations.  In both cases Jupiter and Saturn are locked in 2:1 MMR, with starting semimajor axes of 5.25 AU and $\sim$8.33 AU. In the JSREG simulation (left panels) the gas giants' initial eccentricities are 0.025 and their orbits exhibit regular motion.  In the JSCHA simulation (right) the giants' initial eccentricities are 0.03 and their orbits are chaotic. Their mutual orbital inclination is initially 0.5 degrees. Each system was integrated for 136 Myr, and to perform the Fourier analysis we used an output timestep of 2 years; the top panels show the power spectra for $\varpi_{Sat}$ for each case. The middle and bottom panels show the dynamical excitation of the main belt in the two simulations, using a snapshot at 40 Myr. Some asteroids in the JSCHA simulation have larger eccentricities and inclinations than those observed. This does not mean that our results are inconsistent with the present-day asteroid belt. Highly eccentric and inclined objects are removed from the system during the later evolution of the Solar System: during the dynamical instability between the giant planets \citep{morbidellietal10} and over the subsequent 3.8 billion years \citep{mintonmalhotra10}. These simulations also do not take into account the gravitational influence of the growing terrestrial planets, which may remove a large fraction of dynamically overexcited asteroids. In the JSCHA simulation, asteroids with low orbital inclination  are also observed between 1.8 and  2 AU. Equally, these objects do not exist in the real belt today.  It is highly likely that these objects will be also removed from this region during the accretion of the terrestrial planets  and by the effects of secular resonances when Jupiter and Saturn reach their current orbits (e.g. the $\nu_{16}$ is at $\sim$1.9 AU today; \citep{froeschlescholl89}).}
\end{figure*}

During the gas giants' chaotic evolution the ${\rm \nu_6}$ secular resonance jumps across the entire asteroid belt.  To estimate the radial extent of the jumps of the ${\rm \nu_6}$ we first compute the precession frequency that an asteroid would have with Jupiter and Saturn in the 2:1 resonance. Using linear secular theory (see \cite{murraydermott99}), this frequency is given by:

\begin{equation}
\tiny
{A  = \frac{1}{8\pi}n_{Ast}\left[ \frac{M_{Jup}}{M_{\odot}}\left( \frac{a_{Ast}}{a_{Jup}}\right)^2  b_{3/2}^{(1)}(\alpha_{Jup})  +  \frac{M_{Sat}}{M_{\odot}} \left( \frac{a_{Ast}}{a_{Sat}}\right)^2  b_{3/2}^{(1)}(\alpha_{Sat}) \right],  }
\end{equation}

where ${\rm n_{Ast}}$, ${\rm a_{Ast}}$ are the mean motion and semi-major axis of the asteroid. ${\rm M_{\odot}}$, ${\rm M_{Jup}}$ and ${\rm M_{Sat}}$ are the solar mass, Jupiter's mass and Saturn's mass, respectively. ${\rm a_{Jup}}$ and ${\rm a_{Sat}}$ are Jupiter and Saturn semi-major axes, respectively. ${\rm b_{3/2}^{(1)}}$ is the Laplace coefficient which is computed in function of ${\rm \alpha_{Jup}}$ and ${\rm \alpha_{Sat}}$ which are given by

\begin{center}
$${\alpha_{Jup} =  \frac{a_{Ast}}{a_{Jup}}} $$
$${\alpha_{Sat} =  \frac{a_{Ast}}{a_{Sat}}}$$
\end{center}

If we assume that $\varpi_{Sat}$ precesses with frequencies between $\sim10^{-5}$ and $\sim10^{-4}/year$ the $\nu_6$ location varies from about 1.4 to 3.5 AU (Fig. 5). Asteroids' eccentricities are thus excited across the entire main belt. An analogous process acts to pump asteroids' inclinations. In the JSCHA simulation, although the $\nu_{16}$ is much stronger and wider than in the JSREG simulation, it does not jump across the entire belt.  However, other resonances contribute to exciting the asteroids' inclinations across the entire belt (see Appendix). 

\begin{figure}[h]
\centering
\includegraphics[trim={2.5cm 0cm 2.5cm 0cm},scale=2.59]{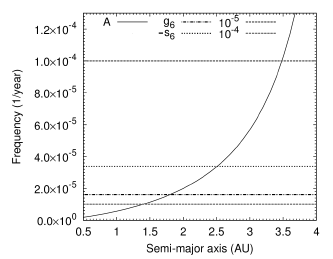}
\caption{Precession frequency of asteroids at different distances from the star calculated from the linear secular theory in dynamical system including Jupiter and Saturn. The gas giants are initially as in the JSREG simulation. The $\nu_6$ and  $\nu_{16}$ locations in the JSREG simulation corresponds to the intersection between the A curve and $g_6$ and $s_6$ frequencies, respectively. Frequencies corresponding to $10^{-5}$ to $10^{-4}$/year are shown to estimate how much the $\nu_6$  jumps across the belt in the JSCHA simulation.}
\end{figure}

The timescale for chaotic excitation of the full asteroid belt is a few million years to hundreds of millions of years depending on the evolution of the gas giants (additional examples in the Appendix). The surviving asteroids broadly match the observed distribution (Fig.~4;  right-hand, middle and bottom panels). In simulations that successfully excited the asteroid belt two conditions were typically observed. First, Jupiter and Saturn's orbits were chaotic in both their eccentricities and inclinations \citep{barnesetal15}. The hardest aspect of the asteroid belt to reproduce is its broad inclination distribution \citep{izidoroetal15b}. Second, Jupiter's eccentricity was not too high. In simulations in which Jupiter's eccentricity remained much larger than its current value of $\sim$0.05 for longer than 100 Myr,  parts of the belt were emptied.  Finally, although we have illustrated this mechanism with Jupiter and Saturn in 2:1 MMR, we observe chaotic excitation in a number of resonant (or near resonant) orbits, including the 2:1, 3:2, 7:4 and 5:3 (see Appendix).

\section{Paths to Chaos in Jupiter and Saturn's early orbits}

We have argued that Jupiter and Saturn's orbits may have evolved chaotically at early times.  In this section we further justify this argument.  We first map the prevalence of chaotic motion in the phase space in the orbits of Jupiter and Saturn.  Next we perform a series of numerical experiments to mimic the orbital migration and resonant capture of Jupiter and Saturn while they were embedded in the gaseous disk.  After the disk's dissipation, a large number of simulations exhibited chaotic behavior.  Finally, we show the long-term dynamical stability between the giant planets evolving  in a chaotic resonant configuration.

\subsection{A map of chaos in Jupiter and Saturn's orbits}

To get a sense of the presence of chaos across the phase space of Jupiter and Saturn's orbital configuration we performed about 9000 simulations to build a dynamical map for a wide range of orbital period ratios between Jupiter and Saturn. The MEGNO (Mean Exponential Growth factor of Nearby Orbits; \cite{cincottasimo00}) chaos indicator is a powerful tool used to identify chaos in dynamical systems.  Chaotic orbits are characterized by a large MEGNO value ($ \left\langle Y \right\rangle \gg~2$) while regular or quasi-period orbits are associated with $ \left\langle Y \right\rangle \rightarrow 2$ (e.g. \citep{cincottasimo00}).

Our initial conditions were set as follows. Jupiter was placed at 5.25 AU while Saturn's semi-major axis was sampled from about 6.25 to 8.6 AU (period ratio between 1.3 and 2.1). Their mutual inclination was sampled randomly from 0 to 0.5 degrees. The eccentricity of Jupiter was randomly selected between 0 and 0.01. Saturn's initial eccentricity ranges from 0 to 0.1. Note that although the initial eccentricity of Jupiter in this set of simulations is smaller than the  corresponding initial values in Figures 1 and 2 this quantity does not remain constant over time. Jupiter and Saturn (secular) interaction leads to eccentricity oscillations which imply that Jupiter's eccentricity may reach values comparable to or even values larger than 0.025-0.03~. Angular orbital elements of both planets were all sampled randomly between 0 and 360 degrees. Simulations were  integrated for 50 Myr using the REBOUND integrator \citep{reintamayo15} computing the MEGNO value of each of these dynamical states.

Figure 6 shows a dynamical map of the behavior of Jupiter and Saturn's orbits at different orbital separations.  The simulations were integrated for fifty million years and the results are color-coded by the MEGNO value.  Black regions are potentially unstable; they show orbital configurations where Jupiter and Saturn undergo close encounters. Orange regions exhibit chaotic motion (with $\left\langle Y \right\rangle \gtrapprox 4$) while blue regions encloses regular motion (with $\left\langle Y \right\rangle = 2$).  Chaotic regions are generally confined in orbital period ratio and are associated with specific mean motion resonances. There is a broad chaotic region near the 2:1 mean motion resonance and narrower regions close to the 5:3 (period ratio of 1.66) and 7:4 (period ratio of 1.75) resonances, with some chaos present just exteriod to the 3:2 resonance (period ratio of 1.5).  

\begin{figure}[h]
\centering
\includegraphics[trim={.25cm 0.cm .7cm .3cm},scale=.722]{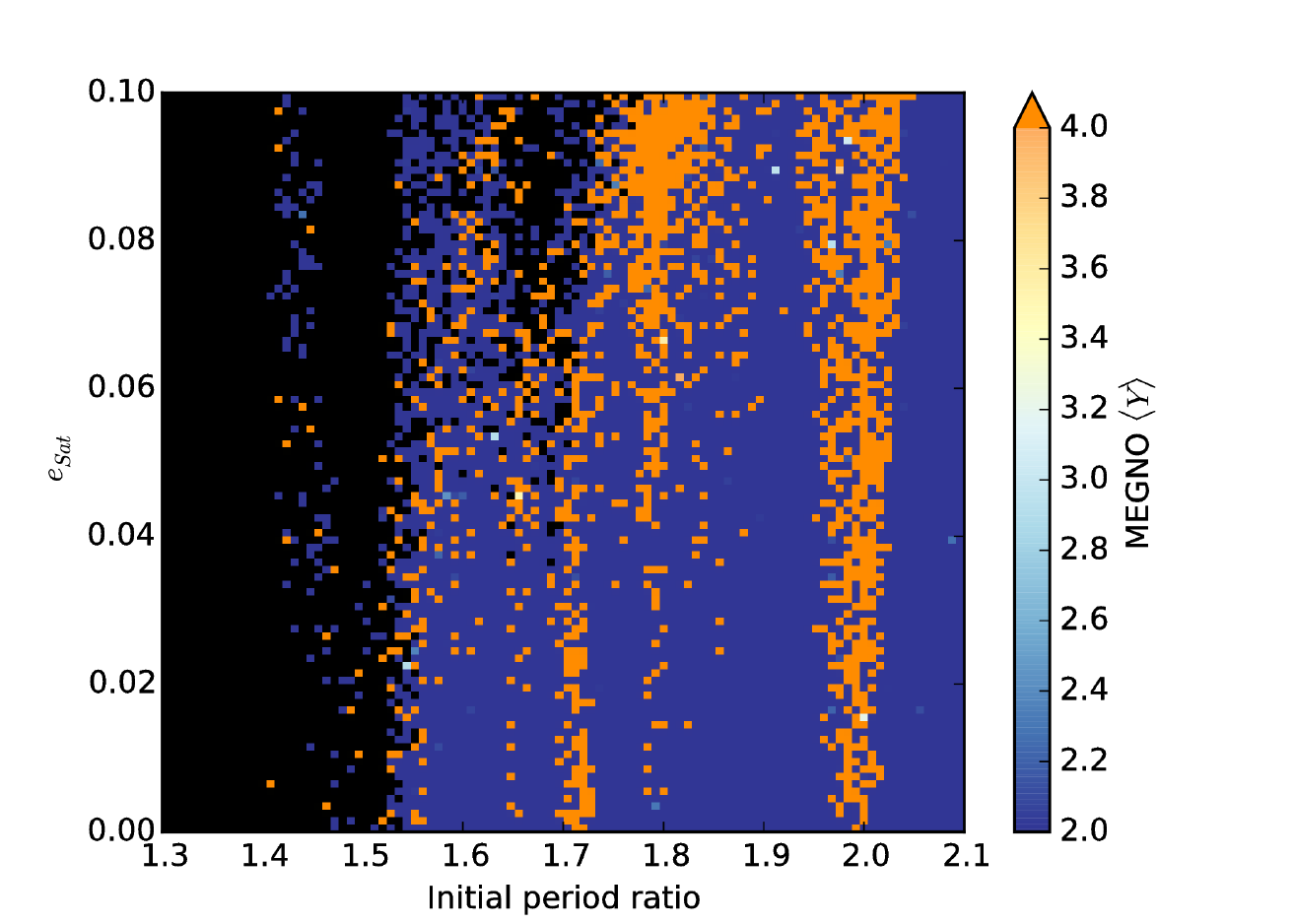}
\caption{A dynamical map of chaos in Jupiter and Saturn's orbits as a function of their initial orbital period ratio. The vertical axes show the initial eccentricity of Saturn (Jupiter's initial eccentricity was randomly chosen between zero and 0.01). The horizontal axes show the initial period ratio between Jupiter and Saturn. Each dynamical state is color-coded showing the MEGNO value after 50 Myr of integration. The black color is used to denote orbits where Jupiter and Saturn mutual distance get smaller than 1~AU. Because we use a symplectic integrator, these orbits would not be solved properly in this case. Blue-ish colors show regular or quasi-periodic motion while orange-ish colors show chaotic orbits.}
\end{figure}

The dynamical map in Figure 6 shows that chaos is common in the phase space available for Jupiter and Saturn's orbits.  Yet their actual orbits at early times were not chosen at random. Rather, the gas giants' orbital configuration was generated by interactions between the growing planets and the gaseous protoplanetary disk, in particular by a combination of orbital migration\cite{linpapaloizou86,ward97a} and eccentricity and inclination damping (e.g., \cite{papaloizoularwood00,bitschetal13}). The dynamical evolution of Jupiter and Saturn in the gas disk depends on the disk properties \cite{massetsnellgrove01,morbidellicrida07,zhangzhou10,pierensraymond11}. To study chaos in Jupiter and Saturn's orbits in the context of orbital migration we perform  N-body simulations using artificial forces to mimic the effects of the gas and also pure hydrodynamical simulations. We present these results next.

\subsection{Chaos in Jupiter and Saturn's orbits in the context of orbital migration}

In our simulations of Jupiter and Saturn migrating in a gaseous disk Jupiter was initially at 5.25 AU and Saturn was placed initially exterior to the 2:1 mean motion resonance with Jupiter (beyond 8.33 AU). Following \cite{baruteauetal14} Jupiter was assumed to migrate in a type-II mode with migration timescale computed by
\begin{equation}
\footnotesize
{ 
t_{m,Jup} = \frac{2a_{Jup}^2}{3\nu} \times min\left( 1,\frac{M_{disk}}{m_{Jup}}\right) 
}, 
\end{equation}
where ${a_{Jup}}$ and ${ m_{Jup}}$ are Jupiter's semi-major axis and mass, respectively. ${\nu}$ is the gas viscosity and ${M_{disk}}$ is the gas disk mass. We modeled the disk viscosity using the  standard ``alpha'' prescription given by ${ \nu = \alpha c_s H}$ \citep{shakurasunyaev73}, where ${c_s}$ is the isothermal sound speed and H is the disk scale height. In our simulations ${\alpha = 0.002}$ and the disk aspect ratio is h$\sim$0.07. To account for the damping of eccentricity and inclination on Jupiter's orbit we assume the following relationship between migration timescale and eccentricity/inclination damping \cite{cridaetal08} 
\begin{equation}
{ t_{e,Jup} = t_{i,Jup} =   t_{m,Jup}  / K}.
\end{equation}
In our simulations, we generally adopted the typical value of $K=10$, although we also performed simulations with $K=1$ and $K=100$.

To mimic the migration of Saturn in the gas disk for simplicity we use the type-I migration/damping approach \cite{tanakaetal02,tanakaward04,papaloizoularwood00,cresswellnelson06,cresswellnelson08}. Since Saturn's gap is not fully open this is an acceptable approximation. We stress that our goal here is only to have convergent and smooth migration of Jupiter and Saturn such we can access the plausibility of chaos origin in this kind of simulation. The initial gas surface density at the location of Saturn is  ${ \Sigma_{Sat} = 900 r_{Sat}^{-0.5}~~g/cm^2}$. To implement migration, eccentricity and inclination damping on Saturn, we use the following formulas:

\begin{equation}
\footnotesize
{t_{m,Sat} = \frac{2}{2.7 + 1.1\beta} \left(\frac{M_{\odot}}{m}\right)  \left(\frac{M_{\odot}}{\Sigma_g {a}^2}\right)\left(\frac{h}{r}\right)^2 \left(\frac{1+ \left(\frac{er}{1.3h}\right)^5}{1-\left(\frac{er}{1.1h}\right)^4}\right)\Omega_k^{-1}}, 
\end{equation}

\begin{equation}
\scriptsize
{ t_{e,Sat} = \frac{t_{wave}}{0.780} \left(1-0.14\left(\frac{e}{h/r}\right)^2 + 0.06\left(\frac{e}{h/r}\right)^3    + 0.18\left(\frac{e}{h/r}\right)\left(\frac{i}{h/r}\right)^2\right),}
\end{equation}
and
\begin{equation}
\scriptsize
{ t_{i,Sat} = \frac{t_{wave}}{0.544} \left(1-0.3\left(\frac{i}{h/r}\right)^2 + 0.24\left(\frac{i}{h/r}\right)^3    + 0.14\left(\frac{e}{h/r}\right)^2\left(\frac{i}{h/r}\right)\right)}
\end{equation}
where
\begin{equation}
{ t_{wave} = \left(\frac{M_{\odot}}{m}\right)  \left(\frac{M_{\odot}}{\Sigma_g a^2}\right)\left(\frac{h}{r}\right)^4 \Omega_k^{-1}}
\end{equation}
and ${ M_{\odot}}$, ${ a}$, ${ m}$, ${ i}$, and ${e}$ are the solar mass, planet's semi-major axis, planet's mass, orbital inclination and eccentricity, respectively. $r$ is the planet's heliocentric distance. ${ \Sigma_g}$  and ${ \beta}$ are the gas disk surface density and gas surface profile index at the planet's location, respectively. In our simulations, in the case of Saturn, ${\beta}$ was calibrated from hydrodynamical simulations. The synthetic accelerations to account for the effects of the gas on the planet were modeled as:

\begin{equation}
{ \bold{a}_{mig} = -\frac{\bold{v}}{t_{m,pla}}}
\end{equation}

\begin{equation}
{ \bold{a}_e = -2\frac{(\bold{v.r})\bold{r}}{r^2 t_{e,pla}}}
\end{equation}

\begin{equation}
{ \bold{a}_i = -\frac{v_z}{t_{i,pla}}\bold{k},}
\end{equation}
where ${\rm \bold{k}}$ is the unit vector in the z-direction. The label ``{\it pla}'', which appears in Eq. 8-10 takes the form  ``{\it Jup}'' or ``{\it Sat}''  to refer to accelerations applied to Jupiter and Saturn, respectively.

To ensure the robustness of this method we compared selected simulations with true hydrodynamical simulations using similar disks setup and they agree in terms of the final period ratio, eccentricities and orbital inclination of Jupiter and Saturn.

In simulations in which the ice giants were also included they were assumed to migrate inward in the type-I mode described above.

We performed 1000 simulations that differed only in the initial surface density of the disk. In each case the gas disk was assumed to dissipate exponentially in 1 Myr ($\tau_{gas}$ = 100 kyr).  After the gas disk dissipation simulations were integrated for another 50 Myr. To compute how chaotic the orbits of the giant planets are in each simulation we again used the MEGNO chaos indicator incorporated in the REBOUND integrator \citep{reintamayo15}.

Figure 7 shows the results of these simulations. The vertical axis shows the relative initial surface density of the disk and the horizontal axis shows the final orbital period ratio between Jupiter and Saturn. Each final dynamical state of our simulation is represented with a circle whose color shows the MEGNO value of the final dynamical state of Jupiter and Saturn. 

\begin{figure}[h]
\centering
\includegraphics[trim={0.3cm 0cm 0cm 0cm},clip,scale=.72]{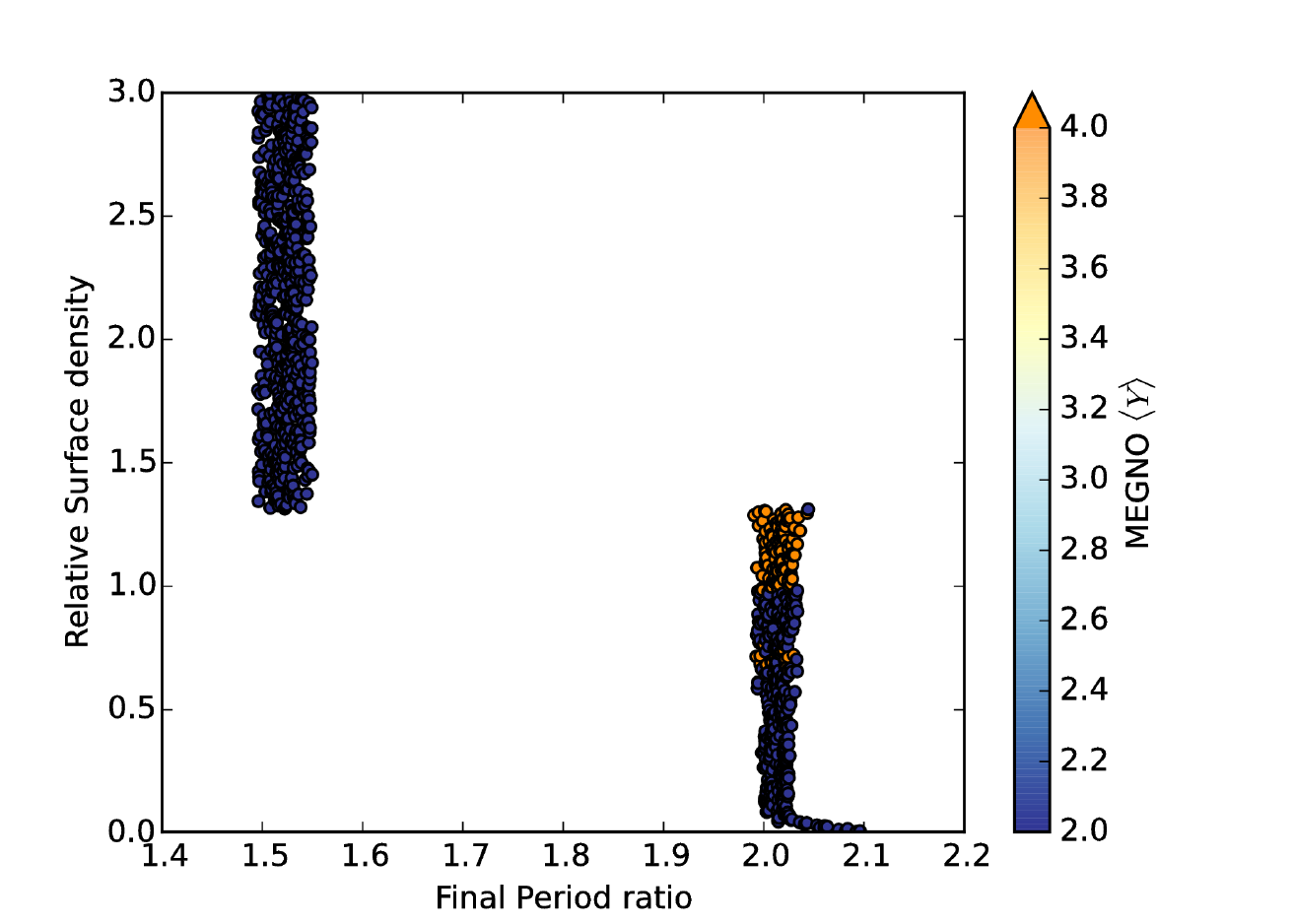}
\caption{Final period ratio of Jupiter and Saturn migrating in N-body simulations including the effects of a 1D gas disk.  The vertical axis shows the relative surface density of the disk. The gas disk lasts 1 Myr. After the gas dissipation the orbits of the giant planets are numerically integrated for another 50 Myr. The color-code shows the MEGNO value of each system at the end of the simulation. In this set of simulations, there is no case where Jupiter and Saturn suffer close-encounters (there is no black points in the figure different from Fig. 6).}
\end{figure}

The experiment presented in Fig 7 produced more regular configurations of Jupiter and Saturn than chaotic ones.  In agreement with pure hydrodynamical simulations \citep{zhangzhou10,pierensetal14}, Jupiter and Saturn typically park in either 3:2 (in high-mass disks) or 2:1 (low-mass disks) mean motion resonance. We did not find chaos in simulations where Jupiter and Saturn ends in 3:2. All our instances of chaos were related to Saturn and Jupiter's period ratio being close to 2. We observe chaos in about 1-25\% of these simulations depending on the disk parameters and K-value (a parameter that defines the ratio between the migration timescale and eccentricity/inclination damping; see Eq. 3). We expect that chaotic configurations for Jupiter and Saturn in 3:2 may be more likely for larger eccentricities than our migration and damping prescriptions allow (see Figure 6).

We performed 15 hydrodynamical simulations using the GENESIS code (Pierens 2005). None of the simulations with Jupiter and Saturn migrating in the disk generated chaos in Jupiter and Saturn's orbits. However, both hydrodynamical simulations and our N-body simulations with synthetic forces are extremely idealized. A number of factors could change the outcome. First, the protoplanetary disks in these simulations are typically laminar and perfectly axi-symmetric. Second, the gas disk's dissipation in numerical simulations is in general poorly modeled with an exponential density decay over the entire disk. Third, at the end of the gas disk phase planets should evolve in a sea of planetesimals and leftover building blocks of their own process of formation. In these simulations we did not include any external perturbation in the system (but see Section 2). Fourth, the migrating ice giants (and/or their building blocks; \cite{izidoroetal15a}) represent a further source of perturbations.

In fact, the results shown in Figures 6 and 7 are complementary in nature. The setup of the simulations in Figure 7 favor deep capture in resonance and regular motion between Jupiter and Saturn while that in Figure 6 (because of the random selection of angles) may preferentially put the dynamical state slightly off the libration center and favor mostly chaotic orbits.  The existence of many chaotic configurations in Figure 6 demonstrates the prevalence of chaos and even if the gas giants' orbits behaved regularly immediately after the dissipation of the gaseous disk, there are a number of processes that could have rendered them chaotic.

\subsection{Long-term dynamical stability of the gas giants in a chaotic configuration} 

The present-day orbits of the giant planets are thought to have been achieved by an instability in the giant planets' orbits that occurred long after the dissipation of the gaseous \citep{tsiganisetal05,morbidellietal07,levisonetal11}.  Our model is entirely consistent with a late ($\sim$500 Myr-later) instability in the giant planets' orbits, whatever the exact configuration of Jupiter and Saturn \citep[see][]{nesvornymorbidelli12,batyginetal12,pierensetal14}. In fact, our model is also consistent with a earlier giant planet instability~\citep{kaibchambers16}, as long as there is a sufficiently long interval during which the belt can be chaotically excited. This interval is roughly longer than 10 Myr in most of the simulations were have run, but can be as short as 2 Myr. To illustrate the possibility of a long-term stability between the giant planets  evolving in chaotic orbits  we performed N-body simulations in which Jupiter, Saturn, Uranus and Neptune migrate in a disk. The prescription for migration used in these simulations is analogous to that explained before in this text. Isothermal type-I migration and damping is also applied for the ice giants. Here,  we used again those damping timescales defined in Eq. 4 to 7 and accelerations given by Eq 8 to 10. Figure 8 shows a long-term stable dynamical evolution of the gas giants in  chaotic orbits. At the end of the gas disk phase Jupiter and Saturn are locked in the 2:1 MMR, Saturn-Uranus in 2:1 MMR, and Uranus-Neptune in 3:2 MMR. The system is dynamically stable for over 500 Myr.

\begin{figure}[h]
\centering
\includegraphics[trim={.0cm 0cm .08cm 0cm},scale=1.93]{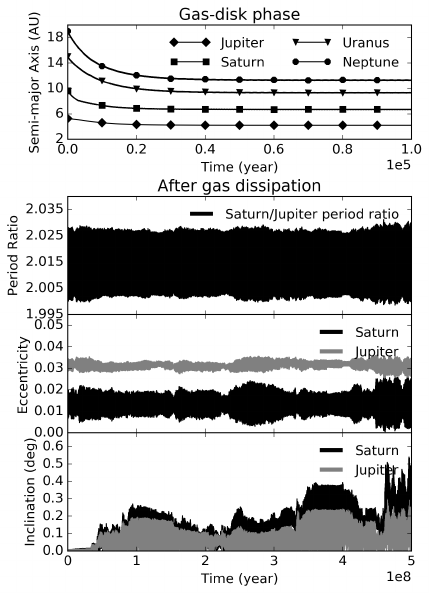}
\caption{{\small An example of the genesis of chaos in the giant planets' orbits and their long term stability.  {\bf Top:} Orbital migration of Jupiter, Saturn, Uranus and Neptune embedded in the gaseous protoplanetary disk.  At the end of the gas disk phase the planets are locked in a chain of mean motion resonances (2:1, 2:1, 3:2). {\bf Bottom:} Long-term evolution of the Saturn-to-Jupiter orbital period ratio, and the gas giants' eccentricities and inclinations over the next 500 Myr. This configuration is consistent with a later instability in the orbits of the giant planets \citep{tsiganisetal05,levisonetal11,nesvornymorbidelli12}.}}
\end{figure}

\section{Discussion}

Chaotic excitation of the asteroid belt represents a novel mechanism for solving a longstanding problem in planetary science.  Yet there are a number of questions that arise when considering whether this mechanism is consistent with our current vision of the Solar System's global evolution.  

In this section we address a number of questions. 

\subsection{Chaotic Excitation vs. the Grand Tack}

 The ``small Mars'' problem highlights the fact that a mass deficit is needed from $\sim 1-4$~AU to explain Mars' small mass relative to Earth's~\citep{wetherill91,raymondetal09}.  In the Grand Tack model, this deficit is generated by Jupiter's long-distance orbital migration from several AU in to 1.5 AU, then back out to beyond 5 AU~\citep{walshetal11,jacobsonmorbidelli14,raymondetal14,brasseretal16}

By explaining the asteroid belt's orbital structure, our results revive models in which the asteroid belt was initially very low mass \citep{izidoroetal15b,levisonetal15}. While standard disk models typically invoke a smooth mass distribution within disks~\citep[for example, the minimum-mass solar nebula model of][generates a smooth disk from discrete planets]{weidenschilling77,hayashi81}, it remains unclear whether the planetary building blocks embedded in these gaseous disks should really follow a smooth distribution.  In fact, models often find that planetesimals preferentially grow in special locations within the disk, such as at pressure bumps~\citep[e.g.][]{johansenetal14}.  Some models naturally create confined rings of planetesimals within broad disks~\citep{survilleetal16}.  There also exist mechanisms to systematically drain solids from certain areas of the disk, thereby creating localized depletions and enhancements~\citep[e.g.][]{moriartyfischer15,levisonetal15,drazkowskaetal16}.  

Our model thus forms the basis of an alternative to the Grand Tack model.  Within the context of this model, Mars' small mass can be explained by a broad mass depletion between Earth and Jupiter's orbits.  The asteroid belt, which could not be stirred by the dynamical effects of local embryos, was chaotically excited by Jupiter and Saturn.

The next step is to search for ways to distinguish between the Grand Tack and this new chaotic model. Tests may be based on observations of small bodies in the Solar System or geochemical measurements. Alternately, the models may be differentiated by more detailed studies of the underlying physical mechanisms involving planetary orbital migration and pebble accretion.

Like the Grand Tack, our model assumes that Jupiter and Saturn migrated during the gas-disk phase into a resonant configuration, mostly likely to the 3:2 or 2:1 MMR \citep{massetsnellgrove01,morbidellicrida07,pierensnelson08,dangelomarzari12,pierensetal14}.  However, the scale of radial migration of Jupiter and Saturn in our scenario could be less specific and much smaller than that in the Grand-Tack scenario, where Jupiter and Saturn migrated inward-then-outward. In our case -- since it is quite unlikely that Jupiter and Saturn grew already in resonance -- only an inward convergent phase of migration between the giant planets is needed to put them into a  resonant configuration.

The Grand Tack and the chaotic excitation model are not contradictory models. In the Grand Tack model, some level of chaotic excitation could have operated between Jupiter and Saturn's two-phase migration and the Nice model instability. However, since the asteroid belt is already sufficiently dynamically excited after the Grand Tack~\citep{deiennoetal16} it would be unnecessary to invoke chaotic excitation.  We also note that both the Grand Tack and chaotic excitation mechanisms may operate with Jupiter and Saturn in either the 3:2 or 2:1 mean motion resonance~\citep[see also][]{pierensetal14}.  Of course, the 3:2 resonance has been much more carefully studied for the Grand Tack, and our results suggest that the 2:1 resonance may be favored with regards to chaotic excitation.  Yet we caution that Jupiter and Saturn's configuration during the disk phase may not necessarily differentiate between the two models.

\subsection{A chaotic young Solar System?}

It is entirely reasonable to imagine a chaotic young Solar System. Several exoplanetary systems with planets on near-resonant orbits are thought to be chaotic, such as 16 Cyg B~\citep{holmanetal97}, GJ876~\citep{riveraetal10}, and Kepler 36~\citep{decketal12}. The present-day Solar System is well known to be chaotic \citep{laskar89,laskar94,sussmanwisdom92}. The orbits of the terrestrial planets undergo chaotic diffusion on a timescale of a few million years \citep{laskar89,batyginetal15}.  It is unknown whether the present-day giant planets are in a chaotic configuration; an accurate determination is precluded by uncertainties in our knowledge of the planets' orbital positions \citep{hayes08,michtchenkoferraz-mello01}. It is often assumed that the early Solar System was characterized by regular motion of the planets \citep{brasseretal13}. Our work suggests that this may not have been the case, and that the structure of the asteroid belt is a signpost that Jupiter and Saturn's early orbits were in fact chaotic.

 We showed that a small nudge could trigger chaos in Jupiter and Saturn's early orbits (Fig. 1). Could another small nudge or the cumulated effects of scattering planetesimals make the system regular again? It is important to note that when scattering of planetesimals takes place not only damping of eccentricity but also radial migration should be observed. This process would lead to divergent migration of Jupiter and Saturn (and ice giants) and not necessarily sinking towards the resonant center and consequently regular orbits. Typically, a late dynamical evolution is envisioned in models of the Solar System evolution (Gomes et al 2005, Levison et al 2011) which implies Jupiter and Saturn hanging out into or very near resonances for hundreds of  Myr after gas disk-phase. What is required for chaotic excitation to work  is simply a sufficient time interval in this phase while the giant planets remain in a chaotic state. This depends also on the chaotic configuration of Jupiter and Saturn and the setup over the disk of planetesimals \citep{nesvornymorbidelli12}.

 Obviously, it would be computationally challenging to perform a systematic analysis to identify all dynamical configurations of Jupiter and Saturn which could excite the belt. However, we indeed found that non-resonant or temporary resonant configurations also can excite the belt. We also recognized that it is not clear why previous classical simulations of terrestrial planet formation did not find similar effects in the belt. One possibility is that the chaotic effects on asteroids have been erased or mitigated by the typical presence of large planetary embryos in the belt in  classical simulations \citep{chambers01,raymondetal06,obrienetal06,izidoroetal14a,izidoroetal15b}. Also, these previous simulations have preferentially considered Jupiter and Saturn near their current orbits or in the 3:2 MMR (in almost circular orbits; see for example \cite{raymondetal09}). In our most successful simulations of belt excitation Jupiter and Saturn have eccentricities of about 0.03-0.05, the latter is consistent with results from hydrodynamical simulations \citep{pierensetal14}.

\subsection{The absence of ``fossilized'' Kirkwood gaps}

The Kirkwood gaps in the asteroid population are created by mean motion resonances with Jupiter.  The most prominent is the gap created by Jupiter's 3:1 mean motion resonance centered at 2.50 AU in the present-day belt. The chaotic excitation of the asteroid belt likely took place before a late instability in the giant planets' orbits (the so-called Nice model; \cite{tsiganisetal05,morbidellietal07,levisonetal11}). If a late instability shifted Jupiter's orbit inward by $\sim$0.2~AU \citep{tsiganisetal05} then there should exist a fossilized gap in the asteroid belt at about 2.6 AU, just exterior to the current Kirkwood gap associated with the 3:1 resonance. No such gap exists (see Section 6.4 of \cite{morbidellietal10}).

We caution that the dataset used to study fossilized gaps (or lack thereof) is relatively sparse, containing just 335 large asteroids (with absolute magnitude $H<9.7$) spread across the entire main belt (see Fig 7 in \cite{morbidellietal10}). The vicinity of the 3:1 Kirkwood gap contains only $\sim$10 asteroids.  A careful statistical analysis using a larger dataset (potentially extending to smaller bodies) would help quantify the depth and width of the missing Kirkwood gaps.

The  belt excitation by the chaotic motion of Jupiter and Saturn seems to require an eccentricity of $\sim$0.03 or more (although we did not perform a systematic analysis on this issue since it is beyond the scope of this paper). It was proposed by \cite{morbidellietal10} and \cite{deiennoetal16} that the pre-instability giant planets must have had almost perfectly circular orbits.  Thus, Jupiter's mean motion resonances were relatively weak and narrow such that the Kirkwood gaps before the instability were virtually nonexistent. Jupiter's eccentricity must remain extremely low ($e_J < 0.01$) to avoid clearing the primordial gap. This requires that during the gas disk phase the orbits of the gas giants were very efficiently damped by the gas \citep{morbidellietal07,pierensetal14}.

The problem  of the fossilized Kirkwood gaps (if it exist al all) appears to have a simple solution.  During the Nice model instability, one or two ice giants are often ejected from the Solar System after suffering close encounters with Jupiter.  To reproduce the Solar System's current architecture, 1-2 additional primordial ice giants may thus have existed before the instability \citep{nesvorny11,batyginetal12}.  During the scattering process a doomed ice giant typically spends time with an orbit interior to (but crossing) Jupiter's.  The scattered ice giant passes through the asteroid belt, often crossing the main belt.  While this interval is short in duration, lasting just a few tens to hundreds of thousands of years, a scattered ice giant perturbs the distribution of asteroids. \cite{brasiletal16} showed that local groupings of asteroids analogous to asteroid families, initially strongly confined in orbital parameter space, are smeared out by the scattered ice giant (see their Figs 4 and 5).  This smearing would fill in any fossilized Kirkwood gaps.  In the context of this evolution of the giant planets, the present-day Kirkwood gaps must have been created after the Nice model instability.

In summary, we expect that any fossilized Kirkwood gaps would have been erased by the Nice model instability \citep{brasiletal16}.  The absence of fossilized gaps cannot be used as a constraint on the giant planets' early orbits.

\subsection{Implantation of C-type asteroids in the Belt}

In this paper we did not address another important characteristic of the asteroid belt: the radial mixing of different taxonomic types of asteroids. The inner part of the main belt is dominated by S-type (water-poor) bodies, while C-type (water-rich) ones are preferentially found in the outer part of the belt, mostly beyond 2.5 AU \citep{gradietedesco82,demeocarry14}. We demonstrate in an upcoming paper that the belt’s chemical dichotomy is a natural,
unavoidable outcome of the gas giants' growth in the gaseous protoplanetary disk. During the gas-accretion phase Jupiter's core (and Saturn as well) destabilizes the orbits of nearby small bodies and implant a fraction of these bodies in the outer asteroid belt \citep[Raymond \& Izidoro, in prep.]{izidoroetal16}. These results together with those already presented here will form the basis for a new model to explain the bulk structure of the asteroid belt.

\section{Conclusions}
We have proposed a new mechanism to explain the puzzling orbital excitation of the asteroid belt. This mechanism requires that Jupiter and Saturn's primordial orbits were chaotic, and we showed that this is indeed a plausible outcome of their growth and migration.  The eccentricities and inclinations of asteroids are excited as a multitude of resonances stochastically jump across the full width of the belt.  This mechanism is consistent with observations and has important implications for our understanding of the early Solar System.  

While this paper explained the orbital distribution of the asteroids, we did not explain another important feature, the taxonomic mixing of the asteroid belt. We reassure the concerned reader that we have a separate novel mechanism to explain this, which will be addressed in an upcoming paper \citep[Raymond \& Izidoro, in prep.]{izidoroetal16}.

\section*{Acknowledgments}
We are very grateful to the referee, Jean-Marc Petit, for his valuable comments that helped to improve an earlier version of this paper. A. I., S. N. R, A. P. and A. M. thank the Agence Nationale pour la Recherche for support via grant ANR-13-BS05-0003 (project MOJO). O. C W. thanks  FAPESP (proc. 2011/08171-3) and CNPq (proc 312813/2013-9) for financial support. Computer time for this study was provided by the computing facilities MCIA (Mésocentre de Calcul Intensif Aquitain) of the Université de Bordeaux and of the Université de Pau et des Pays de l'Adour.  We are also very grateful to the CRIMSON team for managing the mesocentre SIGAMM of OCA where a significant part of these simulations were performed.

\bibliography{library}

\appendix

\section{Additional Simulations}

In this appendix we provide extra details and address one observational constraint. First we present examples of additional resonances that act to chaotically excite the belt in the JSCHA simulation. We then present other examples of the belt's excitation for different chaotic giant planet configurations.

\subsection{The role of different resonances}

 Here we provide more details about the nature of the chaotic excitation mechanism by showing how different chaotically-jumping resonances can excite asteroids' eccentricities and/or inclinations. 
 
Figure 9 show examples of asteroids being excited by the perturbation of chaotic Jupiter and Saturn. These asteroids are in the same simulation as those shown in Figure 3, but simply are in different parts of the main belt.  Different resonances act to increase the eccentricities and inclinations of bodies in the belt.

\begin{figure}[!h]

\centering
\begin{tabular}[b]{@{}p{0.45\textwidth}@{}}
\centering\includegraphics[scale=1.4058]{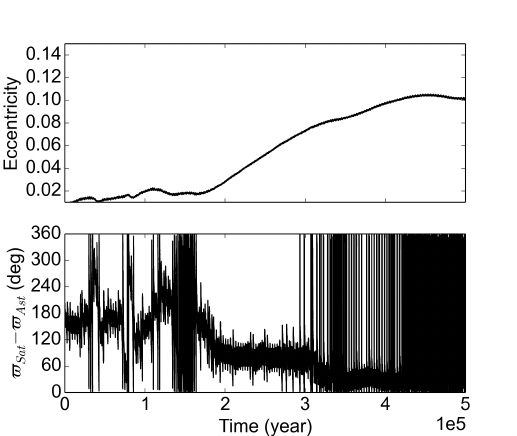} \\
\centering\small (a)  Secular Resonance $\nu_{6}$
\centering\includegraphics[scale=1.4058]{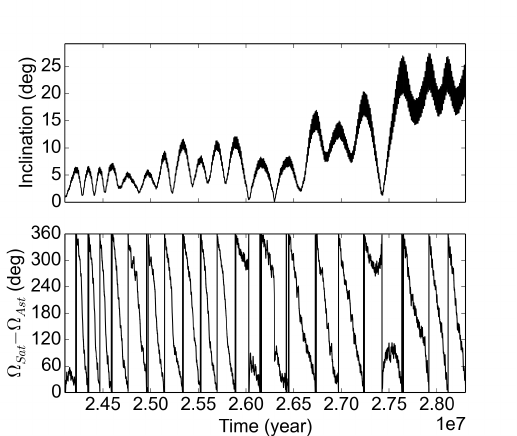}\\
\centering\small (b)  Secular Resonance $\nu_{16}$
\end{tabular}%
\quad
\begin{tabular}[b]{@{}p{0.45\textwidth}@{}}
\centering\includegraphics[scale=1.4058]{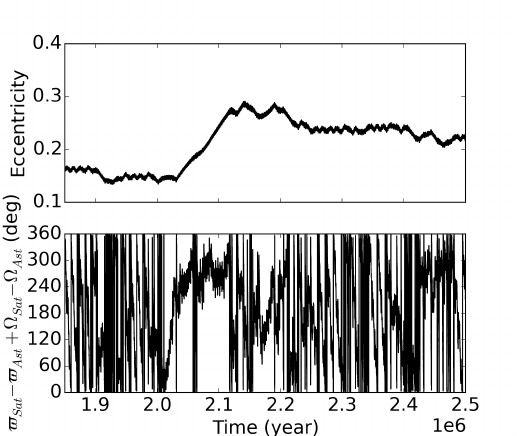}\\
\centering\small (c) Non-linear secular resonance $\nu_{6}$ + $\nu_{16}$
\centering\includegraphics[scale=1.4058]{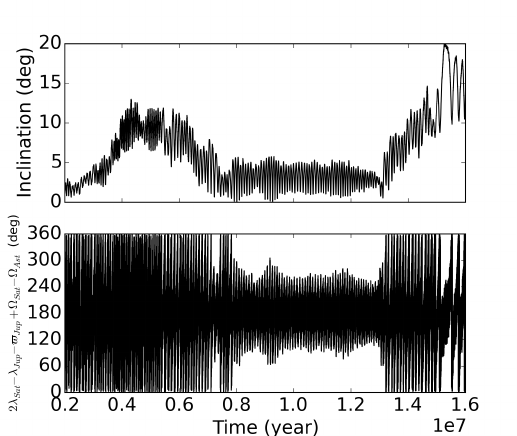} \\
\centering\small (d) Secondary Resonance 
\end{tabular}%
\quad
\begin{tabular}[b]{@{}p{0.45\textwidth}@{}}
\centering\includegraphics[scale=1.4058]{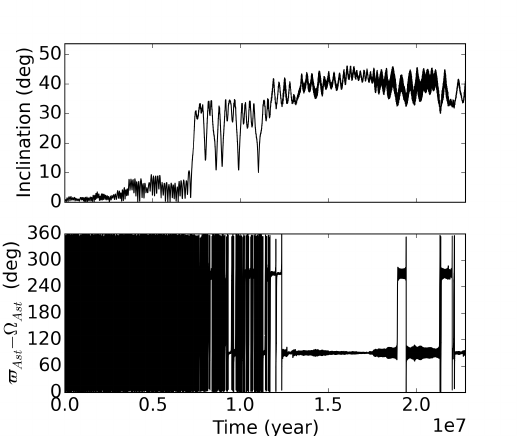} \\ 
\centering\small (e) Kozai Resonance
\end{tabular}%
\quad

\caption{Five examples of dynamical excitation of asteroids in simulations where Jupiter and Saturn have chaotic orbits (simulation JSCHA) caused by different resonances. Asteroids in a) and b) are initially near 2 AU. Asteroids in c), d) and e) reside in the outer part of the main belt, between 2.65 and 3 AU.}
\end{figure}

Figures 10-12 show the Fourier analysis of other angles and comparison between the JSREG and JSCHA simulations. The $\nu_{16}$ secular resonance is much stronger and wider in the JSCHA simulation than the JSREG simulation ($\nu_{16}$ appears at $\sim$2.5~AU; see Figure 4, 5, and 11). However,  its effects in the JSCHA simulation are much more localized than those of the $\nu_6$ in the sense of its power to affect bodies over the whole belt (if bodies have the same proper inclination and eccentricity). We did not perform a systematic analysis of all resonances that contribute to pump inclination of bodies in the whole belt but high order secular and secondary resonances also play an important role for exciting bodies residing away from $\nu_{16}$ in the JSCHA simulation. Perhaps even three body resonances contribute. We stress that the nature of the resonances that contribute to excite bodies in the belt and their strength depends on the chaotic evolution of Jupiter and Saturn.

\begin{figure}[!h]
\centering
\includegraphics[trim={0cm .6cm 0cm 0cm},clip,scale=2.7793]{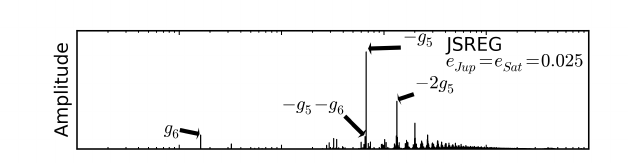}
\includegraphics[trim={0cm 0cm 0cm .3cm},clip,scale=2.7793]{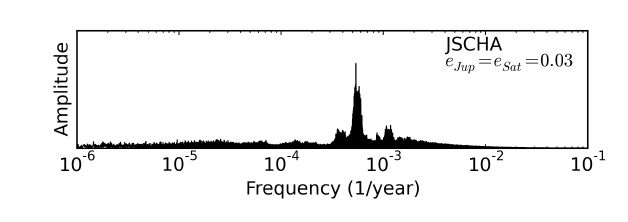}
\caption{Frequency analysis of Jupiter's longitude of pericenter in two simulations in which Jupiter and Saturn are locked in 2:1 mean motion resonance.  {\bf Top:} Both gas giants have initial eccentricities of 0.025 and their orbits exhibit regular motion (simulation JSREG). {\bf Bottom:} The gas giants' initial eccentricities were set to 0.03 and their orbits are chaotic (simulation JSCHA).   Each system was integrated numerically for  136 Myr, and to perform the Fourier analysis we have adopted an output timestep of 2 years. }
\end{figure}

\begin{figure}[!h]
\centering
\includegraphics[trim={0cm .6cm 0cm 0cm},clip,scale=2.7793]{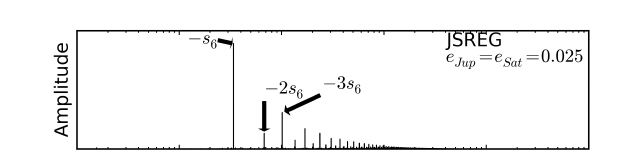}
\includegraphics[trim={0cm 0cm 0cm .3cm},clip,scale=2.7793]{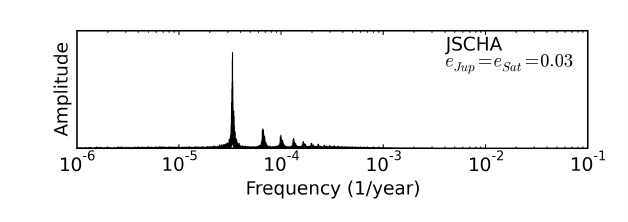}
\caption{Frequency analysis of Jupiter's longitude of the ascending node in two simulations in which Jupiter and Saturn are locked in 2:1 mean motion resonance.  {\bf Top:} Both gas giants have initial eccentricities of 0.025 and their orbits exhibit regular motion (simulation JSREG). {\bf Bottom:} The gas giants' initial eccentricities were set to 0.03 and their orbits are chaotic (simulation JSCHA).   Each system was integrated numerically for  136 Myr, and to perform the Fourier analysis we have adopted an output timestep of 2 years. }
\end{figure}

\begin{figure}[!h]
\centering
\includegraphics[trim={0cm .6cm 0cm 0cm},clip,scale=2.7793]{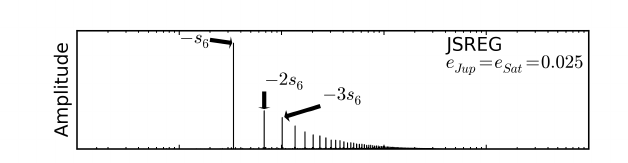}
\includegraphics[trim={0cm  0cm 0cm .3cm},clip,scale=2.7793]{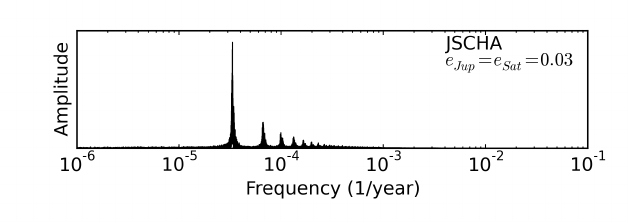}
\caption{Frequency analysis of Saturn's longitude of the ascending node in two simulations in which Jupiter and Saturn are locked in 2:1 mean motion resonance.  {\bf Top:} Both gas giants have initial eccentricities of 0.025 and their orbits exhibit regular motion (simulation JSREG). {\bf Bottom:} The gas giants' initial eccentricities were set to 0.03 and their orbits are chaotic (simulation JSCHA).   Each system was integrated numerically for  136 Myr, and to perform the Fourier analysis we have adopted an output timestep of 2 years.}
\end{figure}

\subsection{Additional Simulations of chaotic excitation of the asteroid belt} 

In this section we present several examples of dynamical excitation of the belt by different orbital configurations between Jupiter and Saturn. In these simulations we used 8000  test particles (in some cases 2000) uniformly distributed between 1.8 and 4.5 AU  We stress that not all our simulations where Jupiter and Saturn had chaotic orbits successfully excited the belt.  Rather, we found a spectrum of outcomes. Some simulations were not able to excite parts of the belt while in other cases the perturbations from Jupiter and Saturn were so strong that the belt was destroyed. We did not perform a systematic analysis looking for the optimal resonant or non-resonant chaotic configuration to excite the belt. However,  we have  used as fiducial case a dynamical configuration where Jupiter and Saturn are in 2:1 MMR.

Figure 13 shows the dynamical evolution of asteroids in a simulation with chaotic Jupiter and Saturn in the 2:1 resonance. Figures 14-17 show the dynamical evolution of asteroids in the belt in simulations where Jupiter and Saturn are in different resonant configurations, with period ratios of $\sim$1.75, $\sim$1.66, $\sim$1.66 and $\sim$1.5, respectively. Simulation were integrated using Symba \citep{duncanetal98} with a 0.1 year timestep.  The total duration of each simulation varied between 20 and 125 Myr; and each simulation's duration is indicated in the bottom panel. Figure 13, 14 and 16 show simulations initially with 8000 test particles. Simulations corresponding to Figures 15 and 17 contain initially 2000 test particles. 

Figure 14 shows a simulation in which the whole belt was excited both in eccentricity and inclination.  However, some regions of the belt were overly depleted compared with others after 20 Myr of integration. 

Figure 15 is a very interesting case. Asteroids in the belt maintained low orbital inclinations and eccentricities for more than 30 Myr. At $\sim$33 Myr a stochastic jump in the positions of the giant planets (consequence of their orbits being chaotic) resulted in a very strong perturbation in the belt (see corresponding panel in Figure 15). Within 2 Myr of the jump, the whole belt was excited to the observed levels of the real one.  This case shows that excitation of the entire belt may be an extremely fast event. However, typically, the complete belt excitation seems to require about $\sim$10~Myr or so.

Figure 16 shows a case where the level of eccentricity excitation is consistent with the observed one. However, the region between 2.1 and 2.8 AU is under-excited in inclination. In general an under-excited inner belt is less of a problem than an under-excited outer belt, because perturbations from the inner parts of the Solar System (e.g., Mars and other remnant planetary embryos) may excite the inner belt but not the outer belt \citep{izidoroetal15b}.  

Figure 17 shows a simulation where Jupiter and Saturn are near the 3:2 resonance. In this case, both eccentricities and orbital inclination of bodies in the belt are modestly under-excited when compared with the real belt. 

Despite some cases failing to reproduce the dynamical excitation of the belt we stress that in all these examples the level of dynamical excitation produced is substantially higher than if Jupiter and Saturn had those respective period ratios but regular orbits. When Jupiter and Saturn have regular orbits only objects near strong secular resonances (as $\nu_6$ and $\nu_{16}$) and mean motion resonances have their eccentricities and inclinations significantly increased relative to the initial value (almost coplanar and circular orbits; \citep{raymondetal09,izidoroetal15b}).

\begin{figure}[!h]
\centering
\includegraphics[trim={0.5cm .9cm .1cm .2cm},clip, scale=1.1]{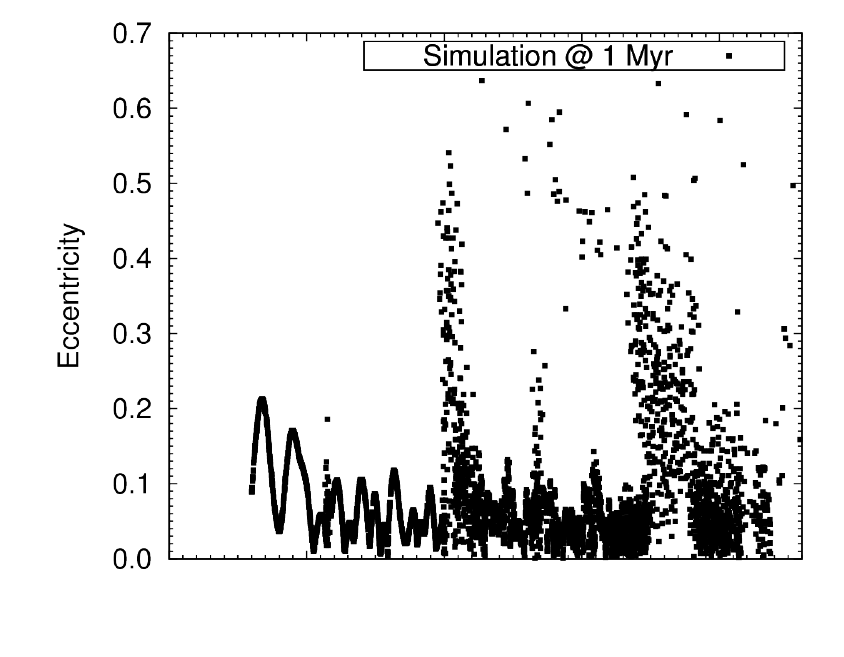}
\includegraphics[trim={0.5cm .9cm .1cm .2cm},clip, scale=1.1]{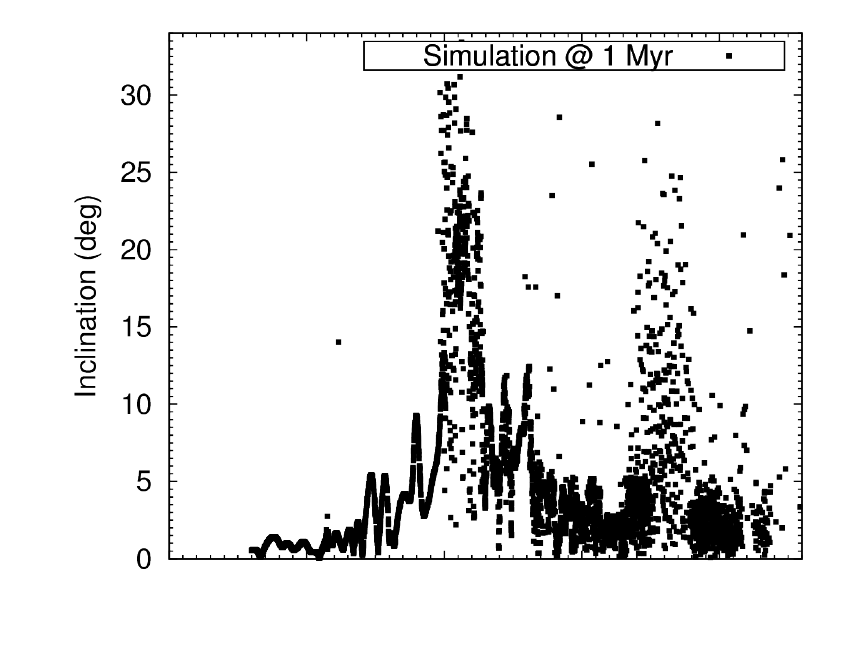}
 
\includegraphics[trim={0.5cm .9cm .1cm .2cm},clip, scale=1.1]{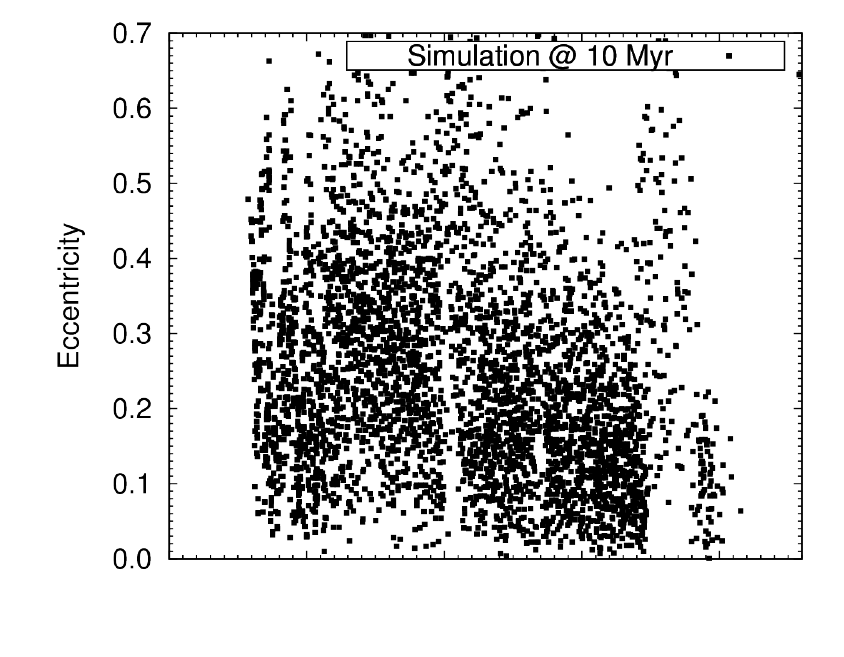}
\includegraphics[trim={0.5cm .9cm .1cm .2cm},clip, scale=1.1]{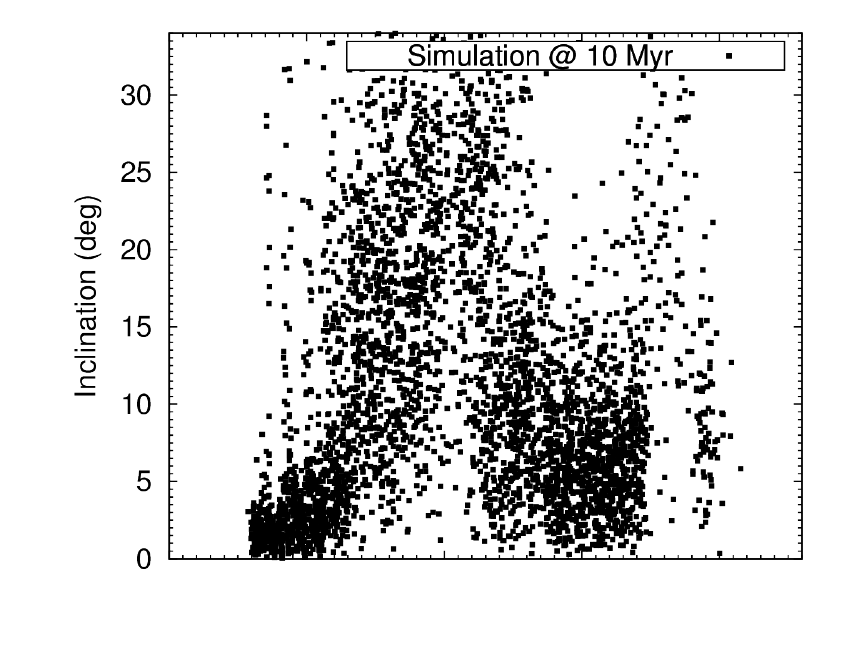}

\includegraphics[trim={0.5cm .0cm .1cm .2cm},clip, scale=1.1]{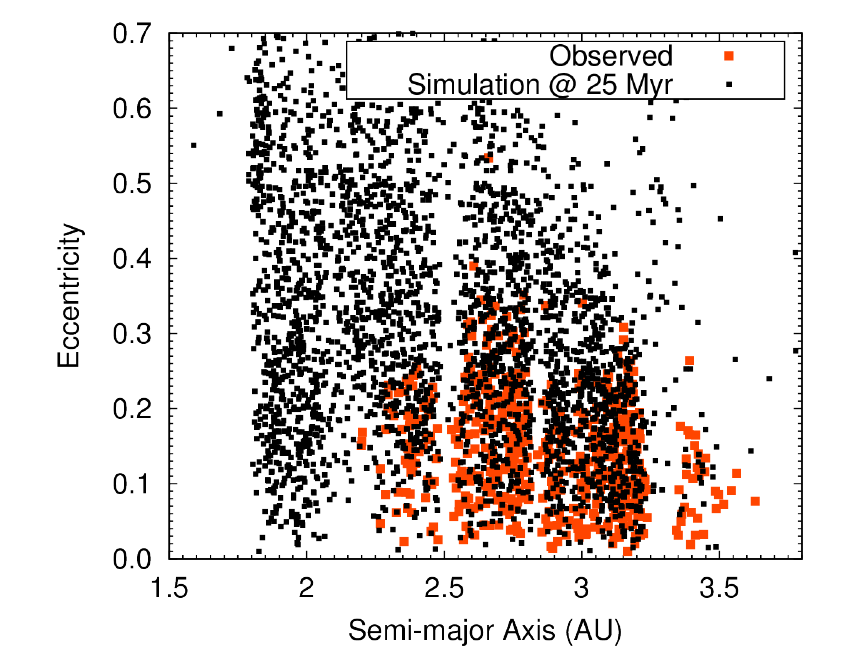}
\includegraphics[trim={0.5cm .0cm .1cm .0cm},clip, scale=1.1]{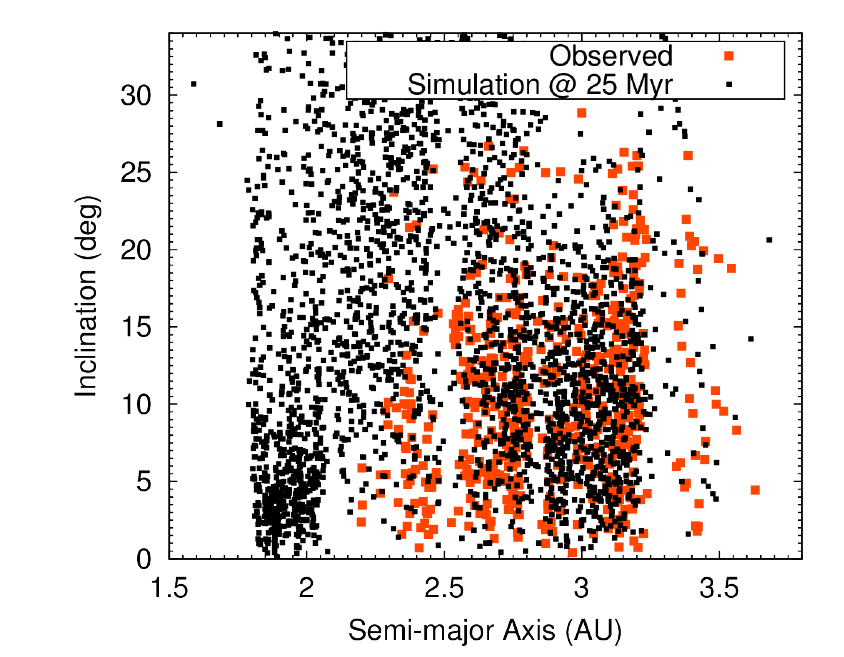}
\caption{Dynamical evolution of asteroids in a simulation with chaotic Jupiter and Saturn in the 2:1 resonance and comparison with real asteroids with diameter larger than 50 km.}
\end{figure}

\begin{figure}[!h]
\centering
\includegraphics[trim={0.5cm .9cm .1cm .2cm},clip, scale=1.1]{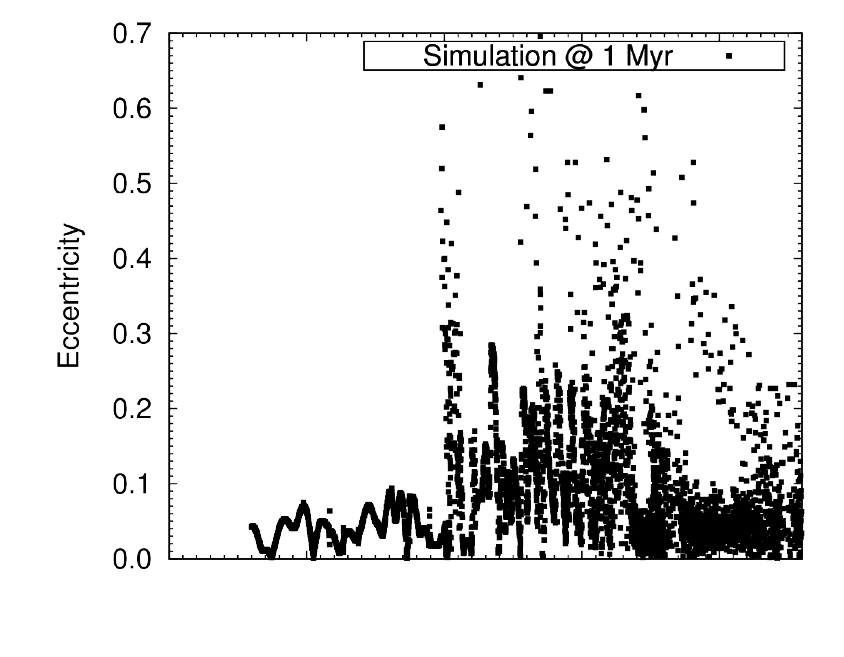}
\includegraphics[trim={0.5cm .9cm .1cm .2cm},clip, scale=1.1]{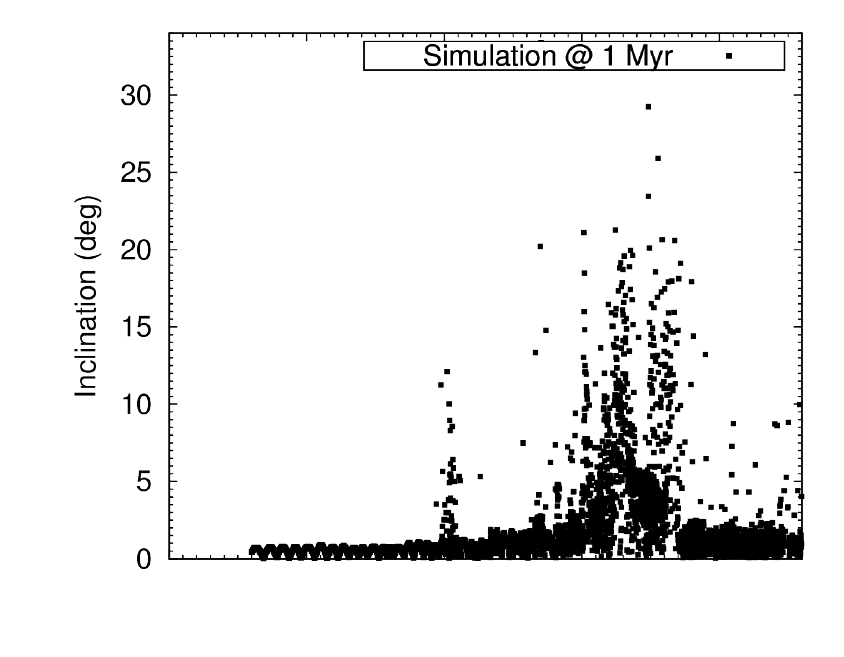}

\includegraphics[trim={0.5cm .9cm .1cm .2cm},clip, scale=1.1]{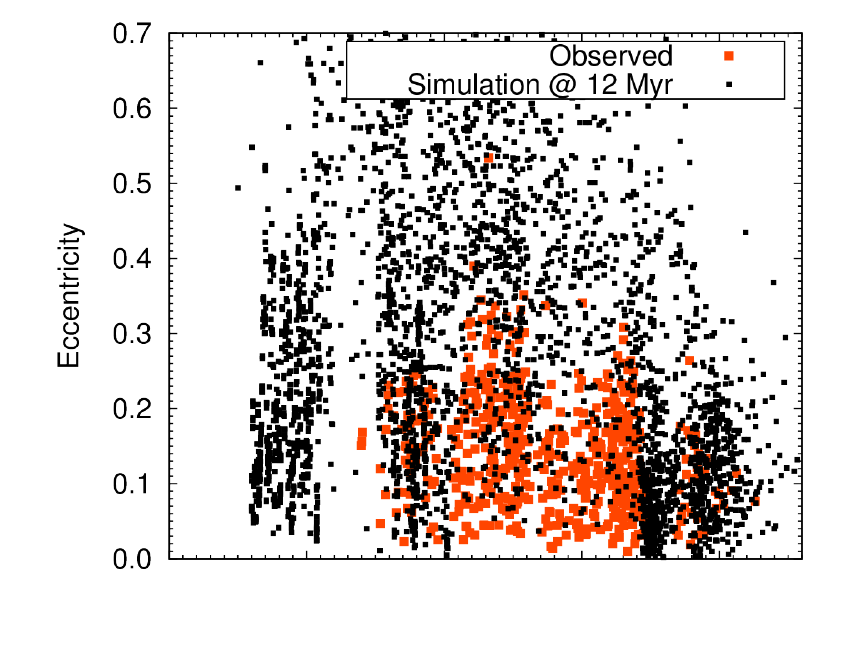}
\includegraphics[trim={0.5cm .9cm .1cm .2cm},clip, scale=1.1]{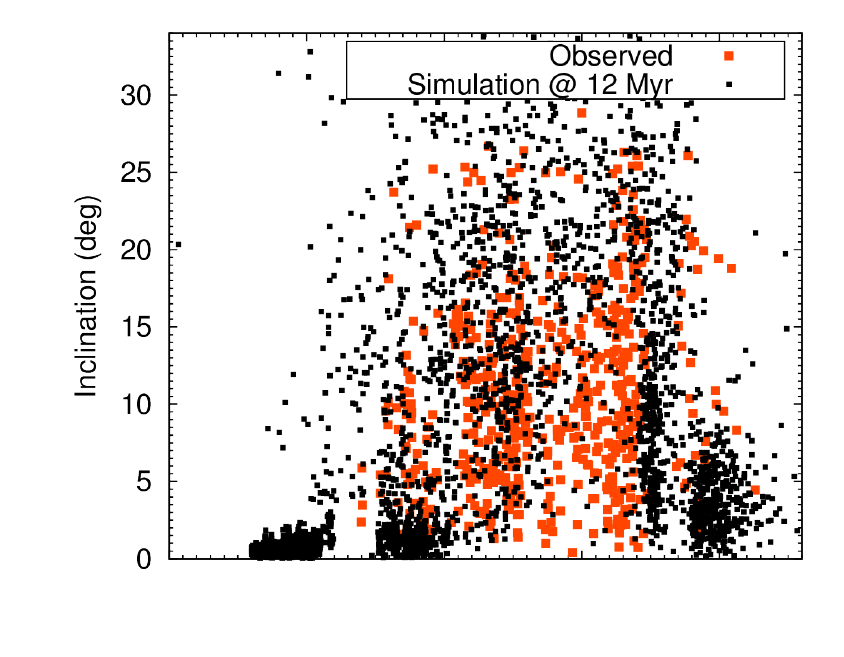}

\includegraphics[trim={0.5cm .0cm .1cm .2cm},clip, scale=1.1]{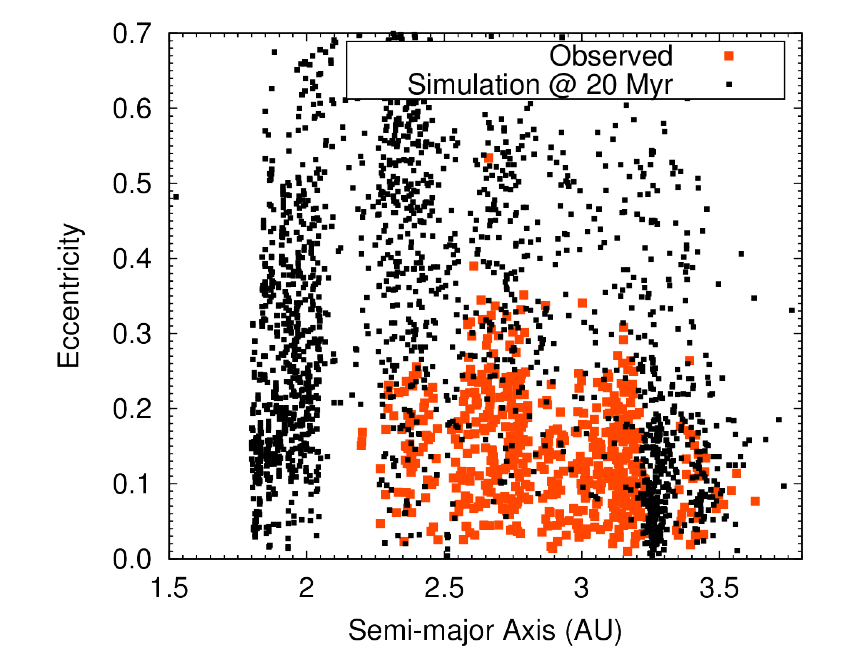}
\includegraphics[trim={0.5cm .0cm .1cm .2cm},clip, scale=1.1]{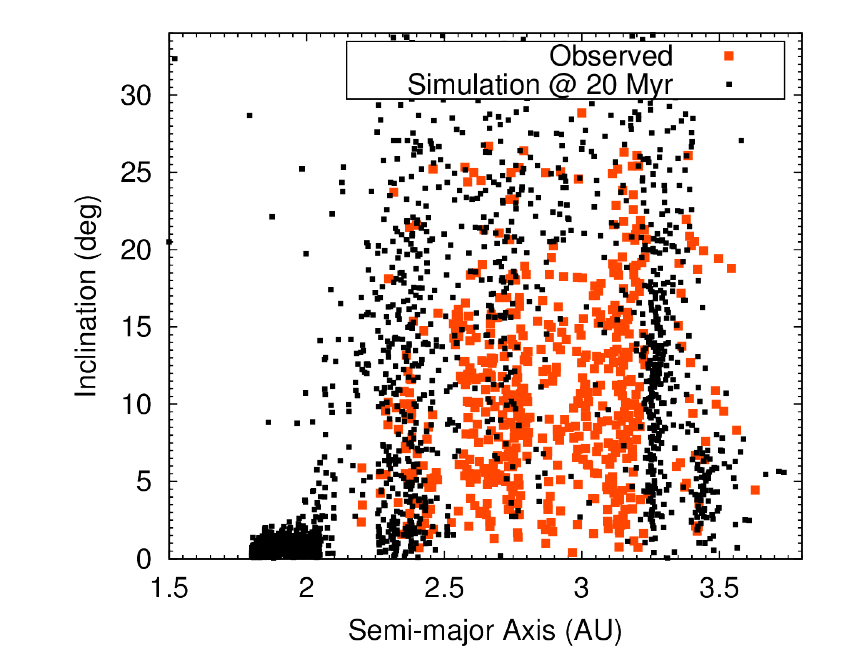}
\caption{Dynamical evolution of asteroids in simulations with chaotic Jupiter and Saturn and comparison with real asteroids with diameter larger than 50 km. The period ratio between Jupiter and Saturn is about 1.75 (one of the resonant angles associated with the 7:4 resonance librate and circulate showing that the planets are near the resonance separatrix).}
\end{figure}

\begin{figure}[!h]
\centering
\includegraphics[trim={0.5cm .9cm .1cm .2cm},clip, scale=.95]{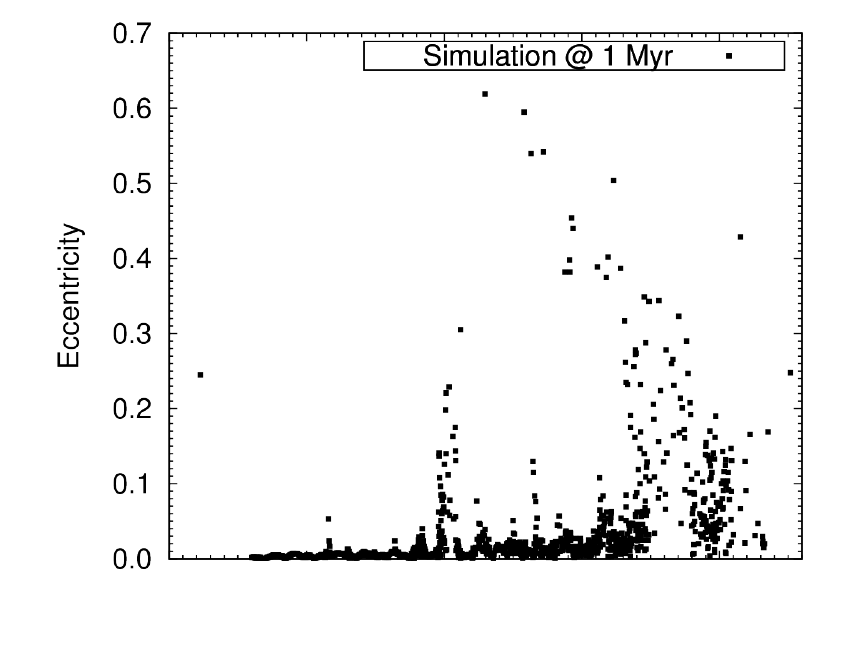}
\includegraphics[trim={0.5cm .9cm .1cm .2cm},clip, scale=.95]{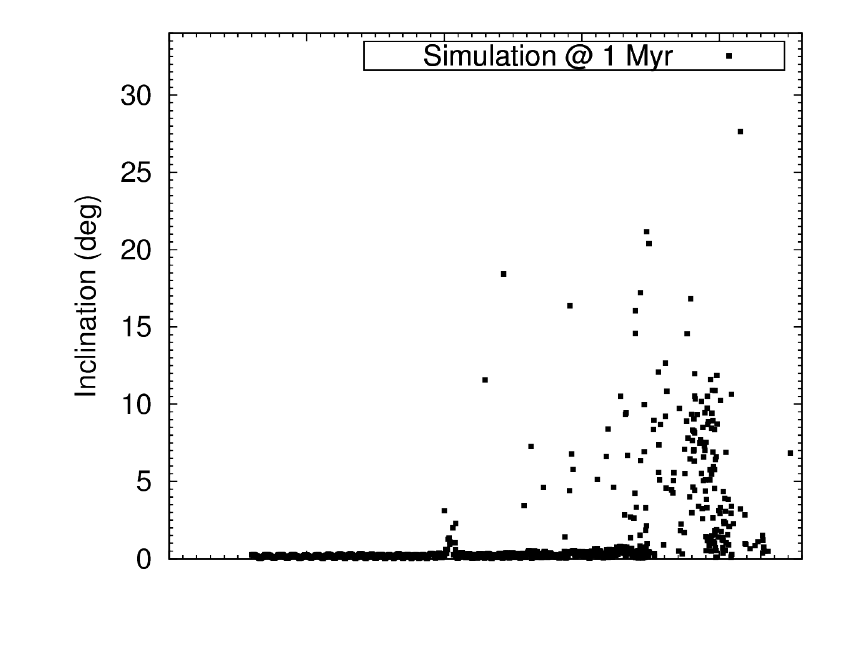}

\includegraphics[trim={0.5cm .9cm .1cm .2cm},clip, scale=.95]{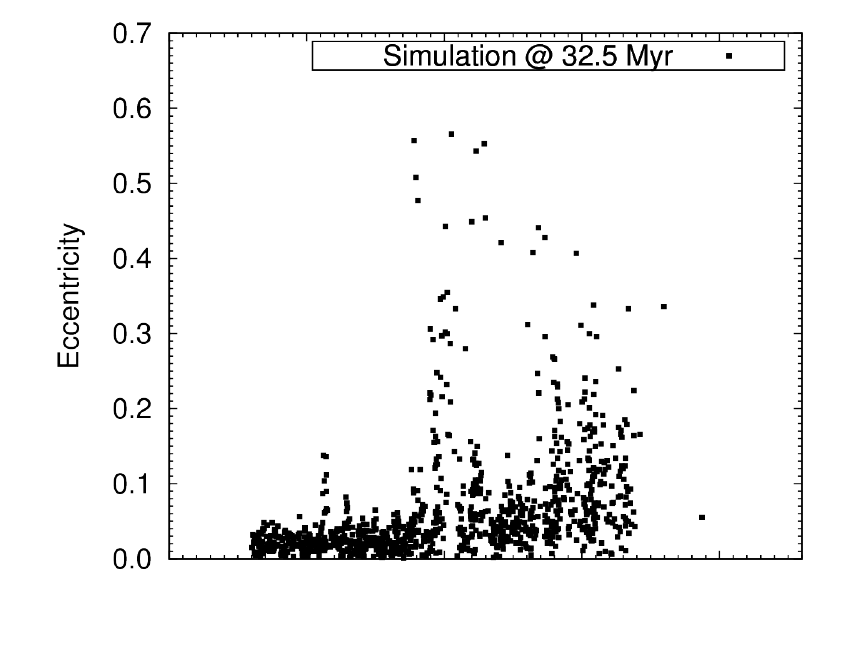}
\includegraphics[trim={0.5cm .9cm .1cm .2cm},clip, scale=.95]{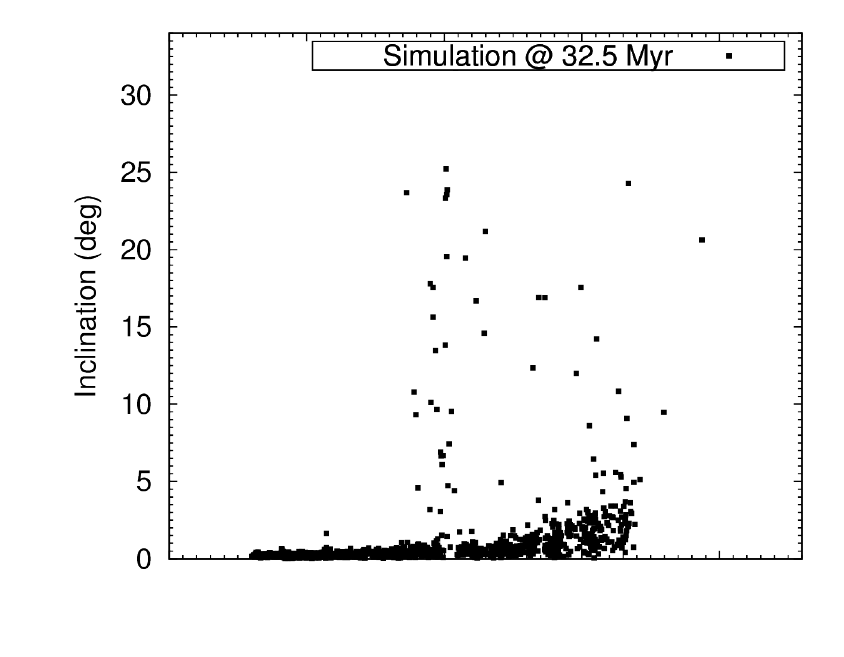}

\includegraphics[trim={0.5cm .9cm .1cm .2cm},clip, scale=.95]{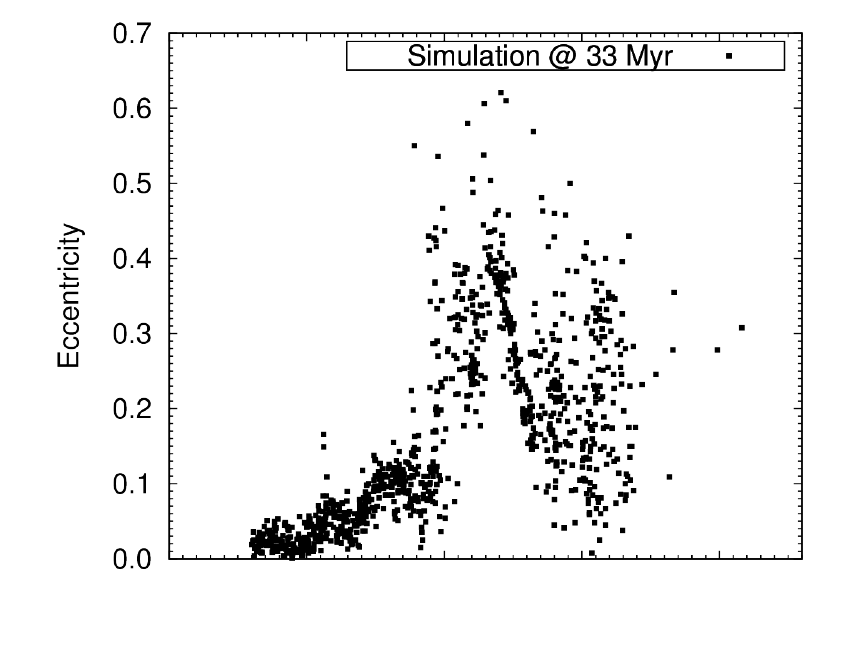}
\includegraphics[trim={0.5cm .9cm .1cm .2cm},clip, scale=.95]{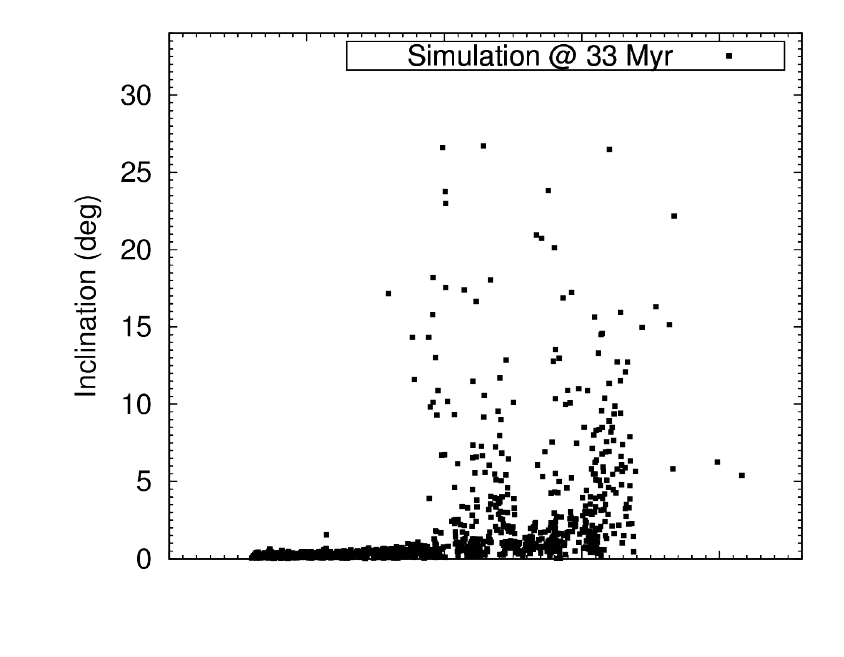}

\includegraphics[trim={0.5cm .0cm .1cm .2cm},clip, scale=.95]{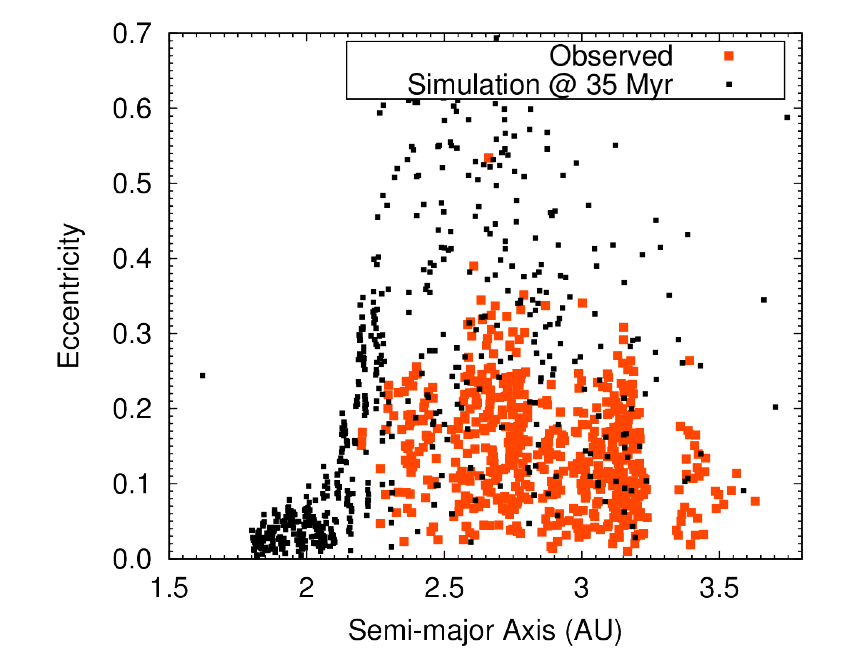}
\includegraphics[trim={0.5cm .0cm .1cm .2cm},clip, scale=.95]{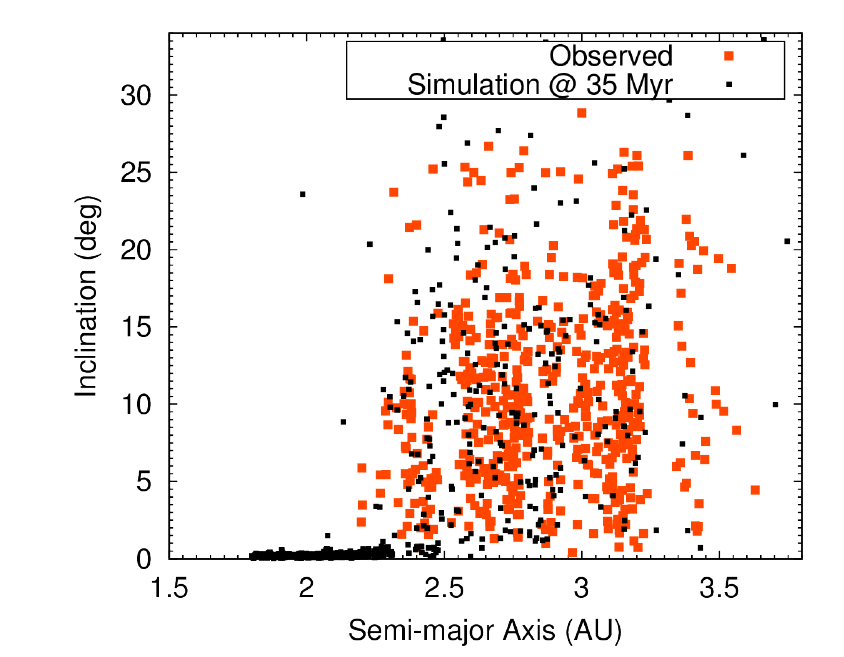}
\caption{Dynamical evolution of asteroids in a  simulation with chaotic Jupiter and Saturn and comparison with real asteroids with diameter larger than 50 km. The period ratio between Jupiter and Saturn is about 1.66 (one of the resonant angles associated with the 5:3 resonance librate and circulate showing that the planets are near the 5:3 resonance separatrix).}
\end{figure}

\begin{figure}[!h]
\centering
\includegraphics[trim={0.5cm .9cm .1cm .2cm},clip, scale=1.1]{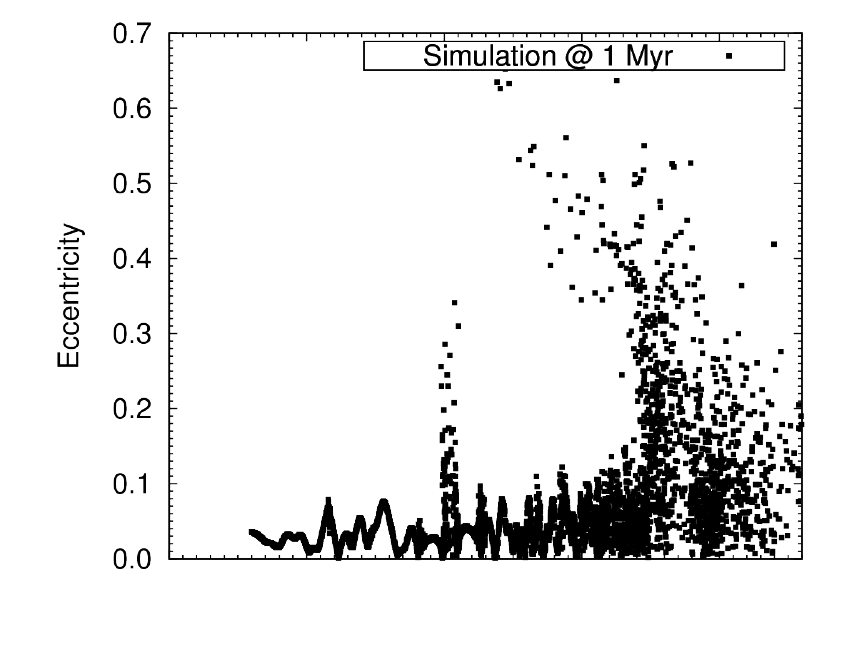}
\includegraphics[trim={0.5cm .9cm .1cm .2cm},clip, scale=1.1]{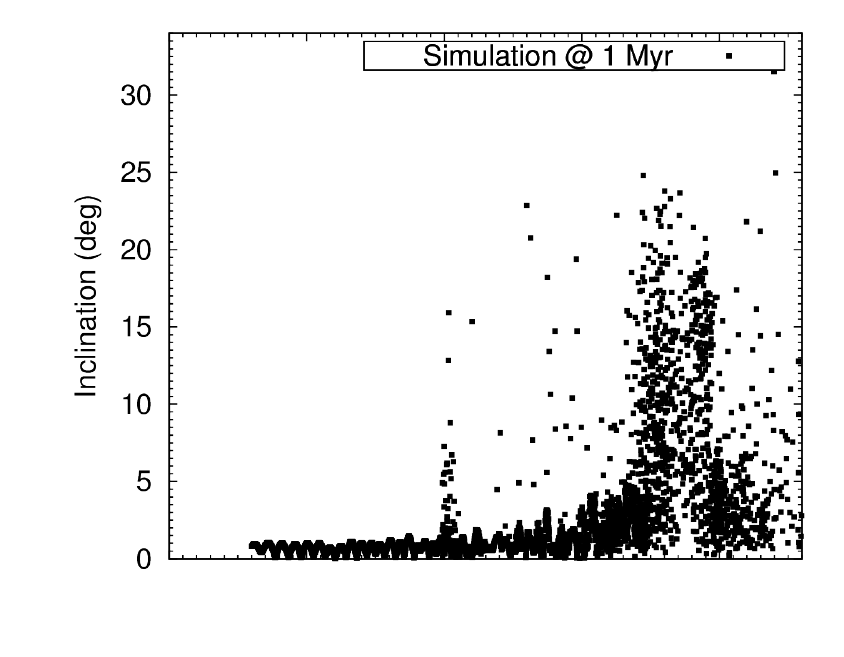}

\includegraphics[trim={0.5cm .9cm .1cm .2cm},clip, scale=1.1]{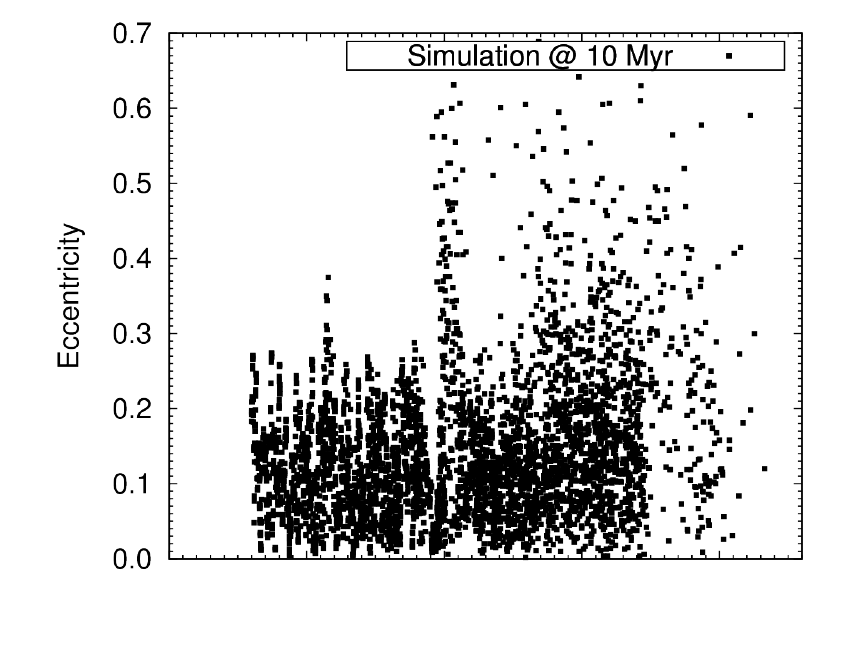}
\includegraphics[trim={0.5cm .9cm .1cm .2cm},clip, scale=1.1]{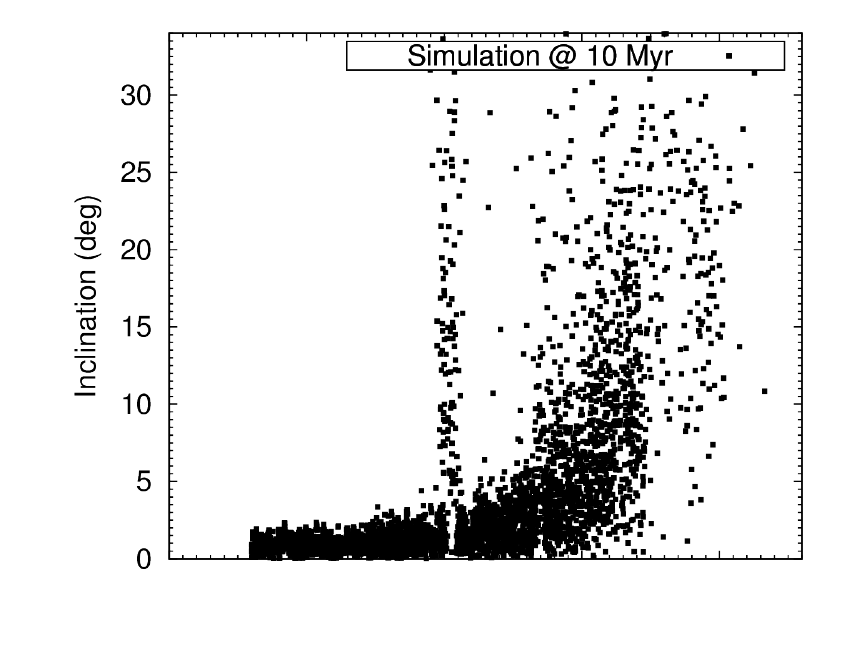}

\includegraphics[trim={0.5cm .0cm .1cm .2cm},clip, scale=1.1]{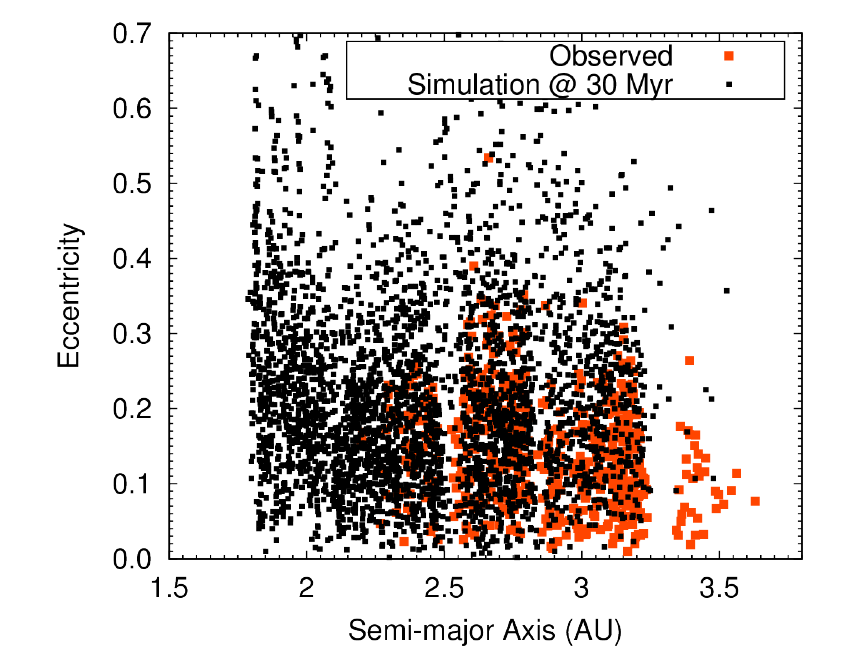}
\includegraphics[trim={0.5cm .0cm .1cm .2cm},clip, scale=1.1]{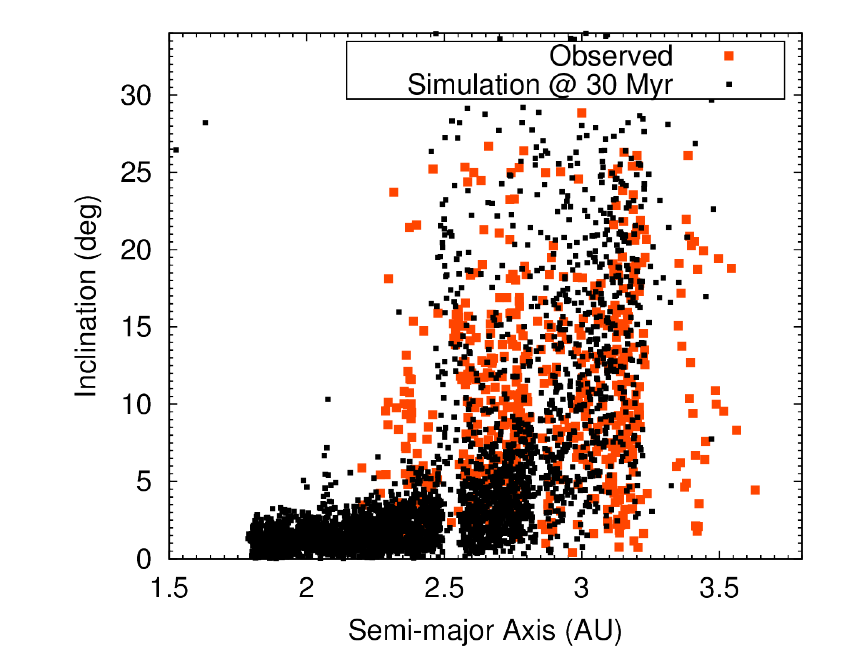}
\caption{Dynamical evolution of asteroids in a simulation with chaotic Jupiter and Saturn and comparison with real asteroids with diameter larger than 50 km. The period ratio between Jupiter and Saturn is about 1.66 (one of the resonant angles associated with the 7:4 resonance librate and circulate showing that the planets are near the resonance separatrix but episodes of libration in the 3:2 resonance are also observed).}
\end{figure}

\begin{figure}[!h]
\centering
\includegraphics[trim={0.5cm .9cm .1cm .2cm},clip, scale=.95]{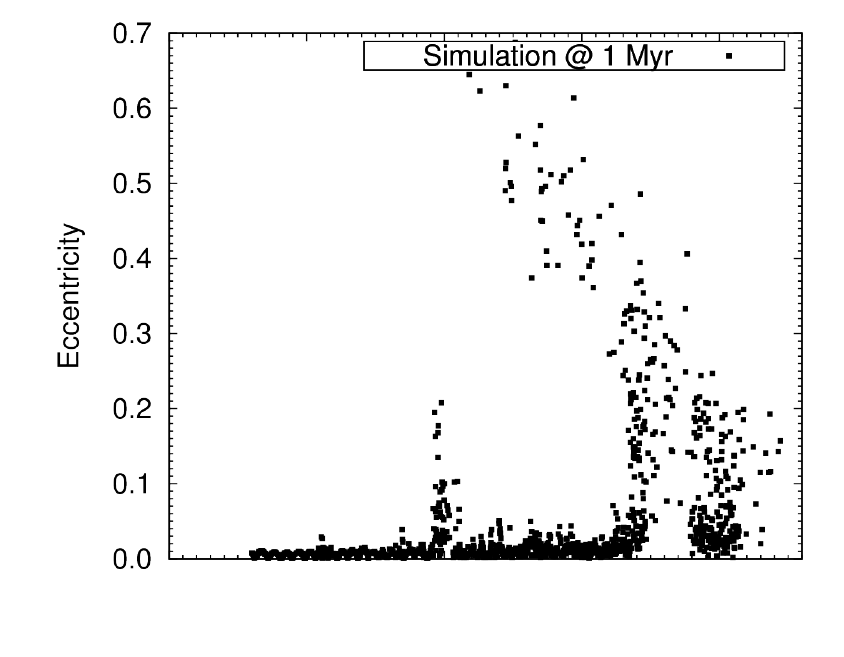}
\includegraphics[trim={0.5cm .9cm .1cm .2cm},clip, scale=.95]{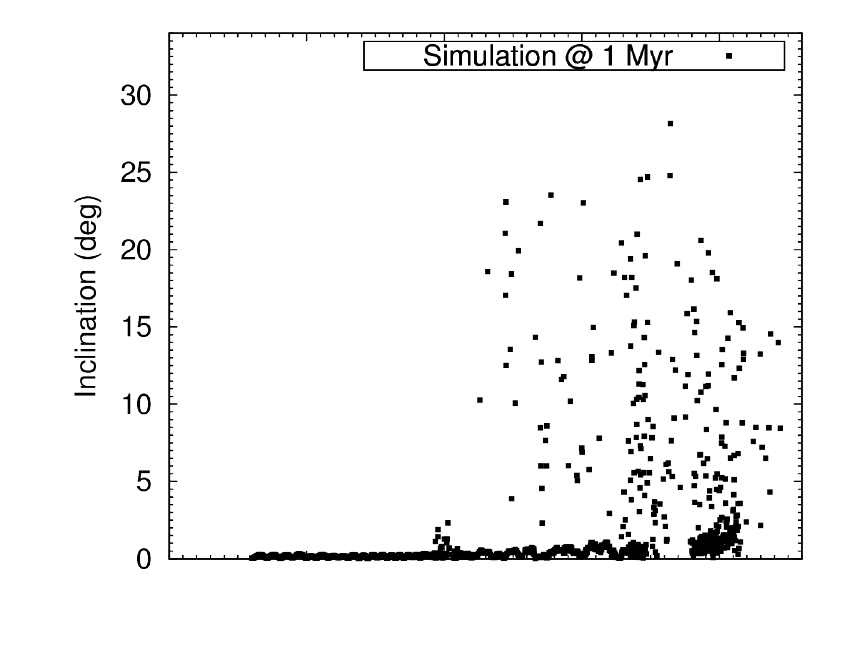}

\includegraphics[trim={0.5cm .9cm .1cm .2cm},clip, scale=.95]{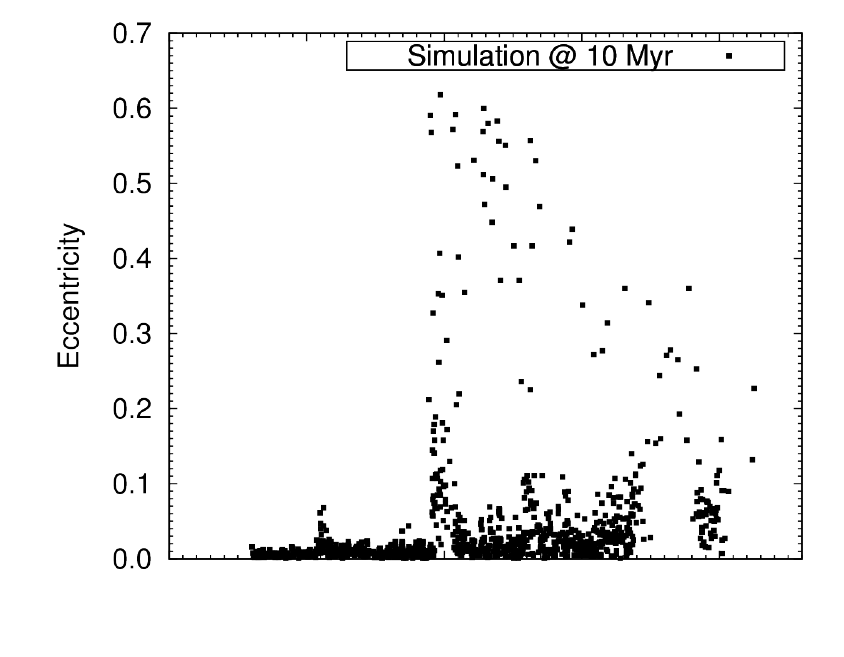}
\includegraphics[trim={0.5cm .9cm .1cm .2cm},clip, scale=.95]{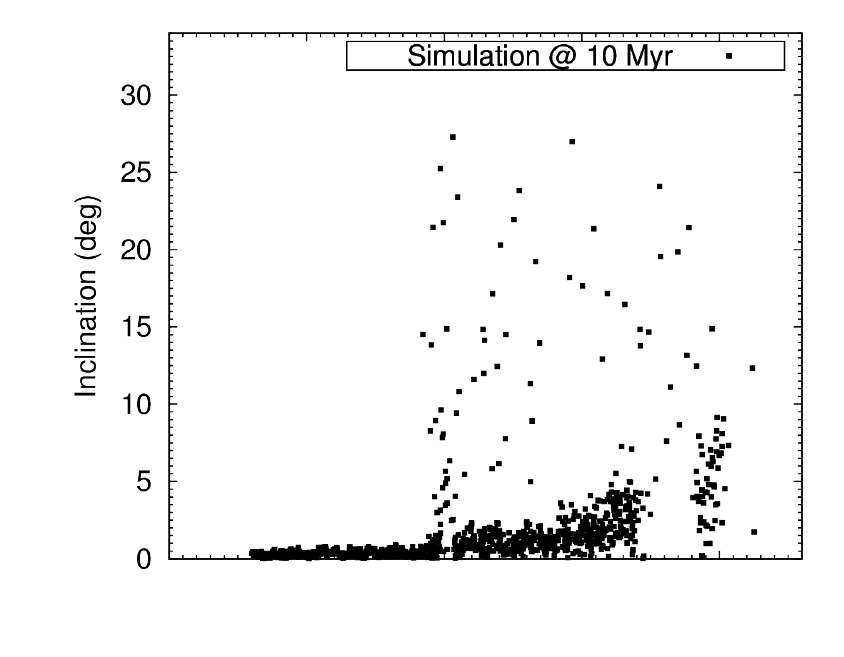}

\includegraphics[trim={0.5cm .9cm .1cm .2cm},clip, scale=.95]{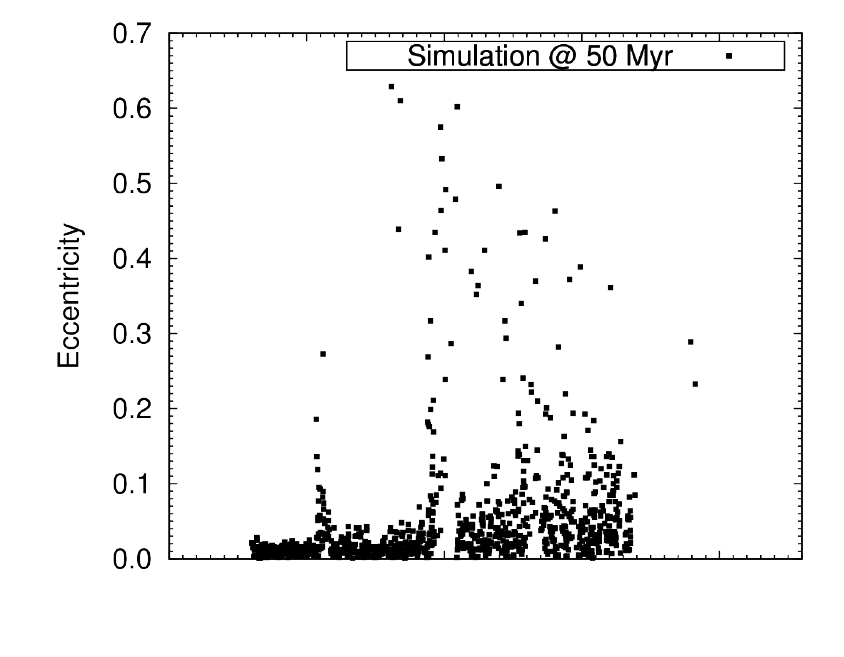}
\includegraphics[trim={0.5cm .9cm .1cm .2cm},clip, scale=.95]{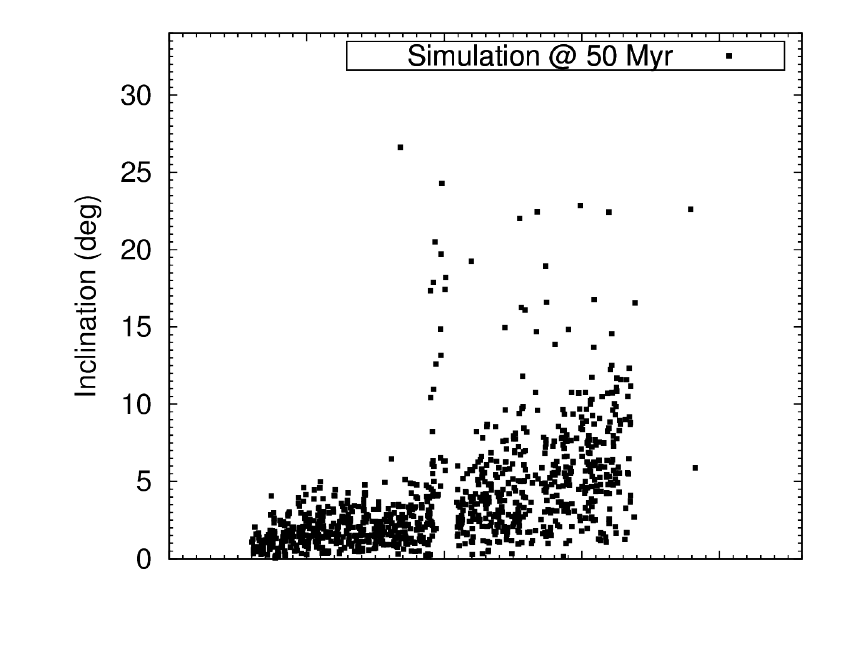}

\includegraphics[trim={0.5cm .0cm .1cm .2cm},clip, scale=.95]{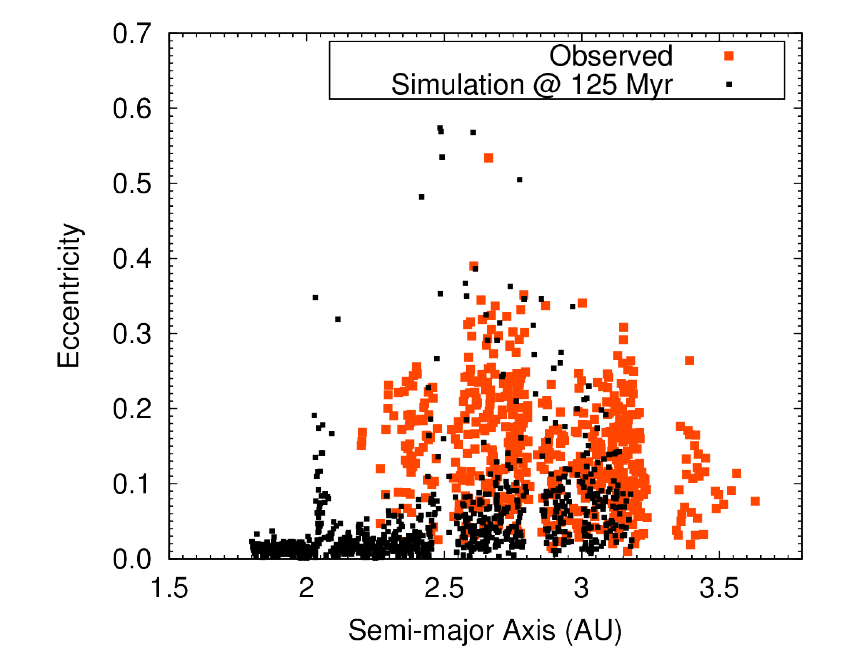}
\includegraphics[trim={0.5cm .0cm .1cm .2cm},clip, scale=.95]{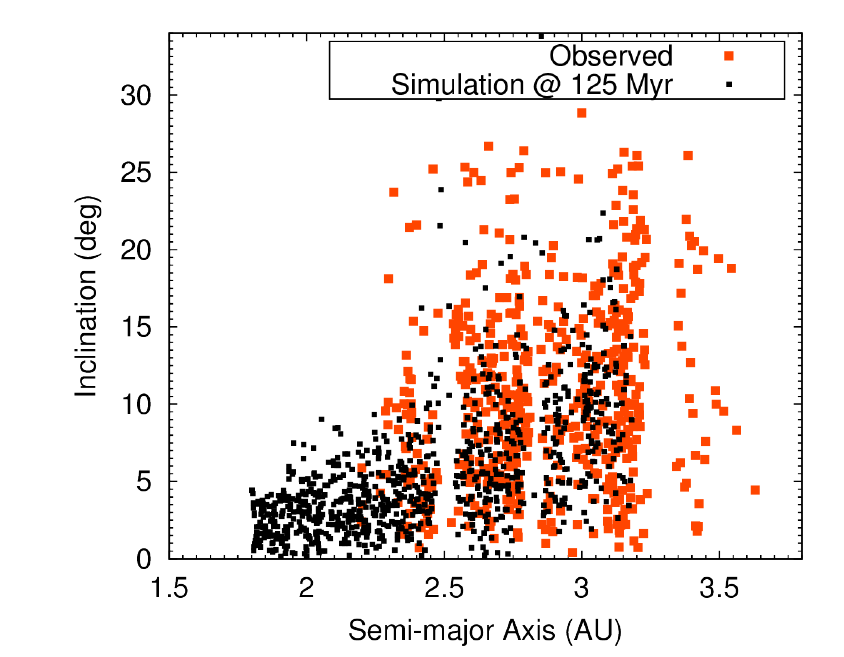}
\caption{Dynamical evolution of asteroids in a simulation with chaotic Jupiter and Saturn and comparison with real asteroids with diameter larger than 50 km. The period ratio between Jupiter and Saturn is about 1.5  (one of the resonant angles associated with the 3:2 resonance librate and circulate showing that the planets are near the resonance separatrix).}
\end{figure}



\subsection{Initial conditions of our main simulations}

 In this section we provide the initial conditions of our simulations corresponding to Figures 1, 2, 4, and 13 to 17.

\begin{table}
\scriptsize
\caption{Initial conditions of our simulations. From left to right the columns show the  corresponding figure in the paper, the simulation, the planet name, the planet mass (solar masses), semi-major axis (AU), eccentricity, orbital inclination (deg.), argument of pericenter (deg.), longitude of the ascending node (deg.) and mean anomaly (deg.).}
\begin{minipage}{\linewidth} 
   \centering 
   \renewcommand{\arraystretch}{0.7}
\begin{tabular}{@{}cccccccccc@{}}
  \hline
  \hline
Figure & Simulation & Planet   & Mass        & Semi-major Axis  (AU) &  Eccentricity   & Inclination (deg.)  &  Argument             & Longitude of the        & Mean \\
       &            &          & (${\rm M_{\odot}}$) &                       &         &                     &  of Pericenter (deg.) & ascending node (deg.)   & Anomaly (deg.) \\ 
  \hline
  \hline 
        &          &           &                      &                         &                 &                     &                 &                  &     \\
Fig. 1  &          &           &                      &                         &                 &                     &                 &                  &     \\
        &          & Jupiter   &   $10^{-3}$          &        5.25             &   0.025         &   0                 &  0              & 0                &  0   \\
        &          & Saturn    & 2.85$\times 10^{-4}$ &        8.33385552283305 &   0.025         &   0.5               &  0              & 0                & 180 \\
        &          & Embryo    & 3.0$\times 10^{-7}$  &        12               &   0.035         &   0                 &  0              & 0                & 250 \\
\\
\hline
        &          &           &                      &                         &                 &                     &                 &                  &     \\
Fig. 2  &          &           &                      &                         &                 &                     &                 &                  &     \\
        & JSREG    & Jupiter   &   $10^{-3}$          &        5.4              &   0.025         &   0                 &  0              & 0                &  0   \\
        & JSREG    & Saturn    & 2.85$\times 10^{-4}$ &        8.5719656        &   0.025         &   0.5               &  0              & 0                & 180 \\
        &          &           &                      &                         &                 &                     &                 &                  &     \\
        &  JSCHA   & Jupiter   &   $10^{-3}$          &        5.4              &   0.03         &   0                 &  0              & 0                &  0  \\
        &  JSCHA   & Saturn    & 2.85$\times 10^{-4}$ &        8.5719656        &   0.03         &   0.5               &  0              & 0                & 180 \\
\\
\hline
        &          &           &                      &                         &                 &                     &                 &                  &     \\
Fig. 4  &          &           &                      &                         &                 &                     &                 &                  &     \\
        & JSREG    & Jupiter   &   $10^{-3}$          &        5.25              &   0.025         &   0                 &  0              & 0                &  0   \\
        & JSREG    & Saturn    & 2.85$\times 10^{-4}$ &        8.33385552283305 &    0.025         &   0.5               &  0              & 0                & 180 \\
        &          &           &                      &                         &                 &                     &                 &                  &     \\
        &  JSCHA   & Jupiter   &   $10^{-3}$          &        5.25              &   0.03         &   0                 &  0              & 0                &  0  \\
        &  JSCHA   & Saturn    & 2.85$\times 10^{-4}$ &        8.33385552283305 &    0.03         &   0.5               &  0              & 0                & 180 \\
\\
\hline
        &  &           &                      &                         &                 &                     &                 &                  &     \\
Fig. 13 &  &           &                      &                         &                 &                     &                 &                  &     \\
        &  & Jupiter   &   $10^{-3}$          &        5.25             &   0.00324129    &   0                 &  0              & 0                &  0   \\
        &  & Saturn    & 2.85$\times 10^{-4}$ &        8.34031296       &   0.03543211    &    1.33660054       &  338.16098022   & 203.47100830     & 178.91156006 \\
\\
\hline
        &  &           &                      &                         &                 &                     &                 &                  &     \\
Fig. 14 &  &           &                      &                         &                 &                     &                 &                  &     \\
        &  & Jupiter   &   $10^{-3}$          &        5.25             &   0.00115474    &   0                 &  0              & 0                &  0   \\
        &  & Saturn    & 2.85$\times 10^{-4}$ &        7.64275074       &   0.04558534   &    1.20436156        &  215.75094604   & 282.47793579     & 155.35372925 \\
\\
\hline
        &  &           &                      &                         &                 &                     &                 &                  &     \\
Fig. 15 &  &           &                      &                         &                 &                     &                 &                  &     \\
        &  & Jupiter   &   $10^{-3}$          &        5.25             &   0.00397692    &   0                 &  0              & 0                &  0   \\
        &  & Saturn    & 2.85$\times 10^{-4}$ &        7.33959627       &   0.01606708    &   0.48118880       &  258.75650024   & 358.17221069     & 160.78182983 \\
\\
\hline
        &  &           &                      &                         &                 &                     &                 &                  &     \\
Fig. 16 &  &           &                      &                         &                 &                     &                 &                  &     \\
        &  & Jupiter   &   $10^{-3}$          &        5.25             &   0.00127964    &   0                 &  0              & 0                &  0   \\
        &  & Saturn    & 2.85$\times 10^{-4}$ &        7.39253807       &   0.02013579    &   1.65362096        &  250.55947876   & 167.58595276     & 31.68255806 \\
\\
\hline
        &  &           &                      &                         &                 &                     &                 &                  &     \\
Fig. 17 &  &           &                      &                         &                 &                     &                 &                  &     \\
        &  & Jupiter   &   $10^{-3}$          &        5.25             &   0.00292543    &   0                 &  0              & 0                &  0   \\
        &  & Saturn    & 2.85$\times 10^{-4}$ &        6.83152723       &   0.02512858    &   0.42025852        &  288.54019165   & 166.67124939     & 291.39468384 \\
\\
\hline
\end{tabular}
\end{minipage}
\end{table}

\end{document}